\documentclass [12pt,a4paper]{article}
\usepackage {amsmath}
\usepackage {graphicx}
\graphicspath{{images/}}
\usepackage {comment}
\usepackage{eqnarray}
\usepackage{relsize}
\usepackage {graphicx}
\graphicspath{{images/}}
\usepackage {comment}
\usepackage{authblk}
\usepackage{cite}
\usepackage{multicol}
\usepackage{longtable}
\usepackage{float}
\restylefloat{table}
\usepackage{multirow}  
\usepackage{amssymb}                    
\newcommand \bprime {\backprime\hspace{-.11em}      }   
\date{}
\title{Differential quadrature element for second strain gradient beam theory}

\author{Md Ishaquddin\thanks{Corresponding author: \textit{E-mail address: ishaq.isro@gmail.com}}, S.Gopalakrishnan\thanks {\textit{E-mail address: krishnan@iisc.ac.in; Phone: +91-80-22932048}}}

\begin{document}

\maketitle
\vspace{-10mm}
\noindent\textit{Department of Aerospace Engineering,Indian Institute of Science 
Bengaluru 560012, India}\\

\begin{abstract}

In this paper, first we present the variational formulation for a second strain gradient Euler-Bernoulli beam theory for the first time. The governing equation and associated classical and non-classical boundary conditions are obtained. Later, we propose a novel and efficient differential quadrature element based on Lagrange interpolation to solve the eight order partial differential equation associated with the second strain gradient Euler-Bernoulli beam theory. The second strain gradient theory has displacement, slope, curvature and triple displacement derivative as degrees of freedom. A generalize scheme is proposed herein to implement these multi-degrees of freedom in a simplified and efficient way. The proposed element is based on the strong form of governing equation and has displacement as the only degree of freedom in the domain, whereas, at the boundaries it has displacement, slope, curvature and triple derivative of displacement. A novel DQ framework is presented to incorporate the classical and non-classical boundary conditions by modifying the conventional weighting coefficients. The accuracy and efficiency of the proposed element is demonstrated through numerical examples on static, free vibration and stability analysis of second strain gradient elastic beams for different boundary conditions and intrinsic length scale values. 

\end{abstract}
\section*{1.0 INTRODUCTION}

The differential quadrature method (DQM) was first introduced by Bellman et al.\cite{Bellman} to solve the linear and non-linear partial differential equations. In this technique, the derivative of the function at a grid point is assumed as a weighted linear sum of the function values at all other gird points in the computational domain, leading to a set of algebraic equations \cite{Shu}-\cite{Wangb}. The computational efficiency and accuracy of differential quadrature method has been established in the literature in comparison with the other numerical methods. However, to circumvent the shortcomings related to the imposing multi-boundary conditions and its applications to the generic structural systems, many improved and efficient versions were proposed in recent years \cite{Bert1}-\cite{Zhong}. A comprehensive survey can be found in review paper by Bert et.al. \cite{review} on various aspects of the differential quadrature methods developed by different authors. Most of the above cited research focussed on developing efficient models for classical beam and plate theories which are governed by fourth order partial differential equations \cite{Karami1}-\cite{Wang4}. Some research addressed the solution methodologies for sixth and eighth order partial differential equations by employing strong from of governing equation in conjunction with the Hermite interpolations \cite{Six}-\cite{Eight}. Recently, authors have proposed two novel differential quadrature elements for the solution of sixth order partial differential equations encountered in non-classical higher order continuum theories\cite{ishaqsf}-\cite{ishaqwf}. These elements are based on strong and weak form of governing equations and employees Lagrange and Hermite interpolations respectively.   

The non-classical gradient elasticity theories are well established in the literature for modelling micro-structural behaviour of materials in contrast to the classical theories\cite{Mindlin1}-\cite{Reddy}. These theories are generalized versions of linear elasticity theories incorporating higher-order terms to account for scale effects. In the non-classical gradient theories, the strain energy depends upon the elastic strain and its gradients\cite{Aifantis1}-\cite{Aifantis3}. The first strain gradient theory which is based on strain tensor and its first gradient generates Cauchy's and double stress tensors. This theory renders sixth order partial differential equation and have been extensively applied to study the static and dynamic behaviour of beams and plates \cite{Besko1b}-\cite{Vardo}. Further, various numerical models have been reported in the literature based on this theory to study the behaviour of beams \cite{Pegios}-\cite{BEM}. The second strain gradient theory, which is an extension of the first gradient theory accounting for second gradient of strain tensor have been used limitedly in the literature for few applications \cite{Mssg}-\cite{Assg}. The research on second srain gradient beams is missing in the literature as per the authors knowledge. In this paper, the variational formulation for the second strain gradient Euler-Bernoulli beam is presented for the first time and the associated classical and non-classical boundary conditions are discussed. Employing the above theoretical basis, we develop a differential quadrature element for the second strain gradient Euler-Bernoulli beam theory which is governed by eighth order partial differential equation. Here we use strong form of the governing equations with Lagrange interpolations as test functions. The present element is the extension of the earlier work by authors for sixth order partial differential equation \cite{ishaqsf}. A novel way to enforce the classical and non-classical boundary conditions in the context of differential quadrature framework is presented. The procedure to compute the  higher order weighting coefficients for the proposed element are explained in detail. The efficiency of the proposed differential quadrature element is demonstrated through numerical examples on bending, free vibration and stability analysis.

\section{Second strain gradient Euler-Bernoulli beam theory} \label{Sg_formulation}

In the present study we consider the simlified second strain gradient micro-elasticity theory with two classical and two non-classical material constants  \cite{Mssg,Assg}. The two classical material coefficients correspond to Lam$e^{'}$ constants and the non-classical ones are of dimension length which are introduced to account for non-local effects. The potential energy density function for a second strain gradient theory is represented as:
\begin{align*}
W&= W(\varepsilon_{ij},\partial_{k}\varepsilon_{ij},\partial_{l}\partial_{k}\varepsilon_{ij}) \tag{1}
\end{align*}

The stress-strain relations for 1-D second strain gradient elastic theory are expressed as \cite{Mssg,Assg}
\begin{align*}
{\tau}&= 2\,\,\mu \,\, \varepsilon + \lambda \,\, {\text{tr}} \varepsilon \,\, \text{I} \\
{{\varsigma}}&= g_{1}^{2} \,\, [2 \,\, \mu \,\, \nabla\varepsilon + \lambda \,\, \nabla (\text{tr}\varepsilon) \,\, \text{I}]
 \\
{\bar{\varsigma}}&= g_{2}^{4} \,\, [2 \,\, \mu \,\, \nabla\nabla\varepsilon + \lambda \,\, \nabla\nabla (\text{tr}\varepsilon) \,\, \text{I}]  \tag{2}
\end{align*}

\noindent where, $\lambda$,$\,$ $\mu$ are Lam$e^{'}$ constants and $g_{1}$, $g_{2}$ are the strain gradient coefficients of dimension length. $\nabla=\frac{\partial}{\partial x}+\frac{\partial}{\partial y}$ is the Laplacian operator and $\text{I}$ is the unit tensor. $\tau$, $\varsigma$ and  $\bar{\varsigma}$ denotes Cauchy, double and triple stress respectively, $\varepsilon$ and ($\text{tr}\,\varepsilon$) are the classical strain and its trace which are expressed in terms of displacement vector $\textit{w}$ as:
\begin{align*}
&{\varepsilon}= \frac{1}{2}(\nabla\textit{w}+\textit{w}\nabla)\,\,, \,\,\quad \text{tr}{\varepsilon}= \nabla\textit{w} \tag{3}
\end{align*}

From the above equations the constitutive relations for a second strain gradient Euler-Bernoulli can be expressed as 
\begin{align*}         
{\tau_{x}}= E\varepsilon_{x}, \quad \varsigma_{x}={g}_{1}^{2}\,\,E\,\varepsilon_{x}^{'}, \quad \bar{\varsigma}_{x}={g}_{2}^{4}\,\,E\,\varepsilon_{x}^{''} \\ \varepsilon_{x}=-z\dfrac{\partial^{2} w(x,t)}{\partial{x}^2}\,\,\,\,\,\,\,\,\,\,\,\,\,\,\,\,\,\,\,\,\,\,\,\,\tag{4} 
\end{align*}

\noindent Based on the above constitutive relations the strain energy is written as
\begin{align*}         
{U}= \frac{1}{2}\int_{0}^{L} EI\big[(w^{''})^{2}+g_{1}^{2}(w^{'''})^{2}+g_{2}^{4}(w^{\backprime\backprime\prime})^{2}\big]dx-\frac{1}{2}\int_{0}^{L} P(w^{'})^{2}dx  \tag{5}
\end{align*}

The potential energy of the applied load is given by
\begin{align*}         
{W}= \int_{0}^{L}q(x)w{dx}+\big[Vw\big]_{0}^{L}-\big[M{w}^{'}\big]_{0}^{L}-\big[\bar{M}{w}^{''}\big]_{0}^{L}-\big[\bar{\bar{M}}{w}^{'''}\big]_{0}^{L}  \tag{6}
\end{align*}

The kinetic energy is given as
\begin{align*}         
{K}= \frac{1}{2}\int_{t_{0}}^{t_{1}}\int_{0}^{L}\rho{A}\dot{w}^{2}{dx}{dt}  \tag{7}
\end{align*}

\noindent  where, $E$, $A$ and $I$ are the Young's modulus, area, moment of inertia, respectively. $q$ and $w(x,t)$ are the transverse load and displacement of the beam. $V$, $M$, $\bar{M}$ and $\bar{\bar{M}}$ are shear force, bending moment, double and triple moment acting on the beam.

Using the The Hamilton's principle\cite{Reddyb}
\begin{align*}         
\delta\int_{t_{0}}^{t_{1}}(U-W-K)\,dt=0   \tag{8}
\end{align*}

\noindent and performing the integration-by-parts. The governing equation of motion for a second strain gradient Euler-Bernoulli beam is obtained as
\begin{align*}   \label{EOM_beam}      
EI(w^{\backprime\backprime\prime}-g_{1}^{2}w^{\backprime\prime\prime}+
g_{2}^{4}w^{\backprime\prime\prime\prime\prime})-q+Pw^{''}+\rho{A}\ddot{w}=0 \tag{9}
\end{align*}

\noindent and the associated boundary conditions are: \\

\noindent \textit{Classical} :     
\begin{align*}  \label{eq:BC_Cl_Beam}     
V&=EI[w^{'''}-g_{1}^{2}w^{\bprime\prime}+g_{2}^{4}w^{\bprime\prime\prime\prime}]=0 \hspace{0.5cm}\text{or}\hspace{0.5cm} w=0,\hspace{0.5cm}\text{at}\,\,x=(0,L)\\
 M&=EI[w^{''}-g_{1}^{2}w^{\prime\bprime\prime}+g_{2}^{4}w^{\bprime\prime\prime}]=0 \hspace{0.5cm}\text{or} \hspace{0.5cm}w^{'}=0,\hspace{0.5cm}\text{at}\,\,x=(0,L)
\tag{10}
\end{align*}

\noindent \textit { Non-classical} :         
\begin{align*}  \label{eq:BC_NCl_Beam}       
\bar{M}&=EI[g_{1}^{2}w^{'''}-g_{2}^{4}w^{\backprime\prime}]=0 \hspace{0.5cm}\text{or}\hspace{0.5cm} w^{''}=0,\,\,\,\text{at}\,\,x=(0,L)		\\
\bar{\bar{M}}&=EI\,{g}_{2}^{4}\,w^{\backprime\backprime\prime}=0 \hspace{2.2cm}\text{or}\hspace{0.5cm} w^{'''}=0,\,\,\,\text{at}\,\,x=(0,L)		\tag{11}
\end{align*} \\
\noindent The list of classical and non-classical boundary conditions employed in the present study for a second strain gradient Euler-Bernoulli beam are as follows \\
 
 \noindent \text{Simply supported} :\\
\noindent \textit{classical} :\,\,$w=M=0$ , \,\,\,\textit{non-classical} : $w^{''}=w^{'''}=0$ \,\,at $x=(0,L)$ \\ 

\noindent \text{Clamped} :\\
\noindent \textit{classical} :\,\,$w=w^{'}=0$ , \,\,\,\textit{non-classical} : $w^{''}=w^{'''}=0$ \,\,at $x=(0,L)$ \\

\noindent \text{Cantilever} :\\
\noindent \textit{classical} :\,\,$w=w^{'}=0$ \hspace{0.1cm} at $x=0$ ,\hspace{0.2cm}$V=M=0$ \hspace{0.1cm}at $x=L$ \,\,\,\,\\
\noindent \textit{non-classical} : $w^{''}=w^{'''}=0$ \hspace{0.1cm} at $x=0$ ,\hspace{0.2cm}$\bar{M}=\bar{\bar{M}}=0$ \hspace{0.1cm}at $x=L$ \,\,\,\,\\

\noindent \text{Propped cantilever} :\\
\noindent \textit{classical} :\,\,$w=w^{'}=0$ \hspace{0.1cm} at $x=0$ ,\hspace{0.1cm}at $x=L$ \,\,\,\,\\
\noindent \textit{non-classical} :\,\,$w^{''}=w^{'''}=0$ \hspace{0.1cm} at $x=0$ ,\hspace{0.2cm}$w^{''}=w^{'''}=0$ \hspace{0.1cm}at  $x=L$ \,\,\,\,\\

\noindent \text{Free-free} :\\
\noindent \textit{classical} :\hspace{0.2cm}$V=M=0$ \hspace{0.1cm}at $x=(0,L)$ \,\,\,\,\\
\noindent \textit{non-classical} :$\bar{M}=\bar{\bar{M}}=0$ \hspace{0.1cm}at $(x=0,L)$ \,\,\,\,

\section{Differential quadrature element for second strain gradient Euler-Bernoulli beam} \label{Lagrange_Beam_section}

The $n$th order derivative of the displacement $w(x,t)$ at location $x_{i}$ for a N-node 1-D beam element is assumed as
\begin{align*}    
w_{i}^{n}(x,t)=\sum_{j=1}^{N} L_{j}^{n}(x)w_{j} \tag{12}
\end{align*}

\noindent $L_{j}(x)$ are the Lagrangian interpolation functions defined as\cite{Wangb,Shu},
\begin{align*}    
L_{j}(x)=\frac{\beta(x)}{\beta(x_{j})}=\prod_{\substack{k=1 \\ (k\neq j)}}^{N}\frac{(x-x_{k})}{(x_{j}-x_{k})}   \tag{13}
\end{align*}

\noindent where \\
$\beta(x)=(x-x_{1})(x-x_{2})\cdots(x-x_{j-1})(x-x_{j+1})\cdots(x-x_{N})$ \\
$\beta(x_{j})=(x_{j}-x_{1})(x_{j}-x_{2})\cdots(x_{j}-x_{j-1})(x_{j}-x_{j+1})\cdots)(x_{j}-x_{N})$ \\

The first order derivative of the above shape functions can be written as
\begin{align*}    
A_{ij}={L}^{'}_{j}(x_{i})\begin{cases}
\mathlarger\prod_{\substack{k=1 \\ (k\neq i,j)}}^{N}(x_{i}-x_{k})/\mathlarger\prod_{\substack{k=1 \\ (k\neq j)}}^{N}=(x_{j}-x_{k})\,\,\,\, (i\neq j)\\ \\ 
\mathlarger{\sum}_{\substack{k=1 \\ (k\neq i)}}^{N}\frac{1}{(x_{i}-x_{k})}
\end{cases}\tag{14}  
\end{align*}

\noindent The higher order conventional weighting coefficients are defined as
\begin{align*}    
B_{ij}=\sum_{k=1}^{N} A_{ik}A_{kj} \,,\quad
C_{ij}=\sum_{k=1}^{N} B_{ik}A_{kj} \,,\quad 
D_{ij}=\sum_{k=1}^{N} B_{ik}B_{kj}\,,\quad(i,j=1,2,...,N)\tag{15}
\end{align*}
Where, $B_{ij}$ , $C_{ij}$ and $D_{ij}$ are weighting coefficients for second, third, and fourth order derivatives, respectively.\\

\begin{figure}[H]
\includegraphics[width=1.0\textwidth]{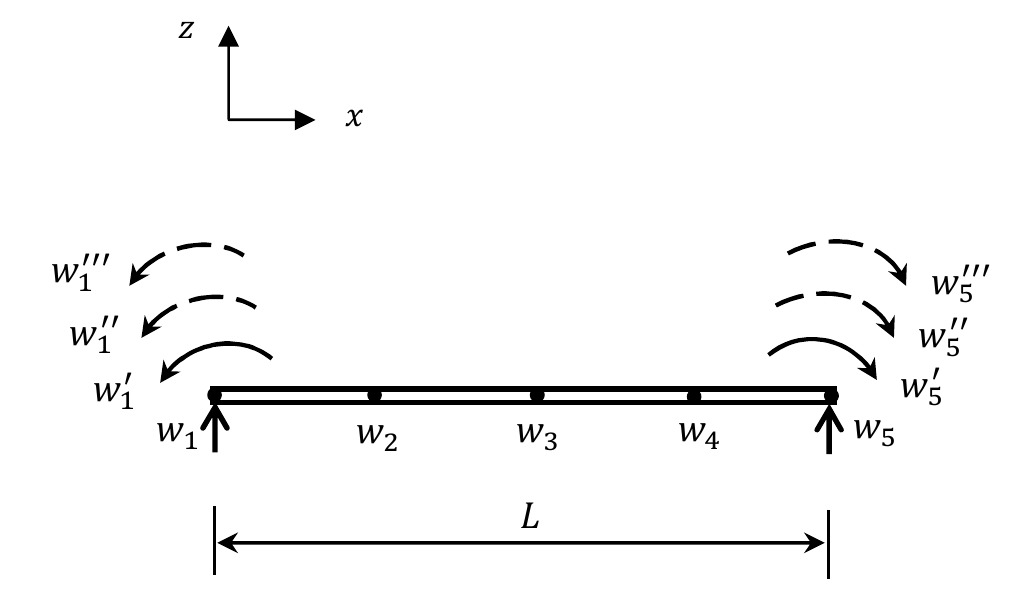}
\centering
\caption{A typical differential quadrature element for second strain gradient Euler-Bernoulli beam with N=5.}
\label{fig:beam}
\end{figure}

\noindent A N-noded second strain gradient Euler-Bernoulli beam element is shown in the Figure \ref{fig:beam}. Each interior node has displacement $w$ as the only degree of freedom (dof), and the boundary nodes has four degrees of freedom $w$, $w^{'}$, $w^{''}$ and $w^{'''}$. These extra boundary degrees of freedom related to higher gradients of displacement are introduced in to the formulation through modifying the conventional weighting coefficients. The new displacement vector now includes the slope, curvature and triple derivative of displacement as additional dofs at the element boundaries as: $\bar{w}=\{w_{1},\cdots w_{N},w^{'}_{1},w^{'}_{N},w^{''}_{1},w^{''}_{N},w^{'''}_{1},w^{'''}_{N}\}$. The modified weighting coefficient matrices accounting for multi-degrees of freedom at the boundaries are derived as follows:\\ 

\noindent \textit{First order derivative matrix}: 
\begin{align*} \label{eq:WC_Beam_Aij}   
\bar{A}_{ij}=\begin{cases}
A_{ij} \,\,\,\, (i,j=1,2,\cdots,N)\\ \\
0 \,\,\,\,\,\,\,(i=1,2,\cdots,N;\,j=N+1,\cdots,N+6)\tag{16}
\end{cases} 
\end{align*}
\noindent \textit{Second order derivative matrix}: 
\begin{align*} \label{eq:WC_Beam_B2ij}
\bar{B}_{ij}=\begin{cases}
B_{ij} \,\,\,\, (i=2,3,\cdots,N-1; \,\,j=1,2,\cdots,N)\\ \\
0 \,\,\,\,\,\,\,(i=2,3,\cdots,N-1; \,\, j=N+1,\cdots,N+6)\tag{17}
\end{cases} 
\end{align*}
\begin{align*} \label{eq:WC_Beam_B2ij}
\bar{B}_{ij}=\sum_{k=2}^{N-1}A_{ik}A_{kj} \,\,\, (i=1,N; \,\, j=1,2,\cdots,N)\tag{18}
\end{align*} 
\begin{align*} \label{eq:WC_Beam_B2ij}
\bar{B}_{i(N+1)}=A_{i1}\,\,; \,\, \, \bar{B}_{i(N+2)}=A_{iN} \,\,\,\,(i=1,N)\tag{19}
\end{align*} 

\noindent \textit{Third order derivative matrix}: 
\begin{align*}\label{eq:WC_Beam_C1ij}
\bar{C}_{ij}=\begin{cases}
\mathlarger{\sum}_{j=1}^{N}\mathlarger\sum_{k=1}^{N}\bar{B}_{ik}A_{kj} \,\,\, (i=2,3,...,N-1) \\ \\ 
0 \,\,\,\,\,\,\,\,\,\, (i=2,3,\cdots,N-1; \,\, j=N+1,\cdots,N+6)\tag{20}
\end{cases}
\end{align*} 
\begin{align*} \label{eq:WC_Beam_C2ij2}
\bar{C}_{ij}=\sum_{k=2}^{N-1}B_{ik}A_{kj}\,\,\  (i=1,N; \,\,j=1,2,\cdots,N)\tag{21}
\end{align*}
\begin{align*} \label{eq:WC_Beam_C2ij22}
\bar{C}_{i(N+3)}=A_{i1}\,\,; \,\, \, \bar{C}_{i(N+4)}=A_{iN} \,\,\,\,(i=1,N)\tag{22}
\end{align*}

\noindent \textit{Fourth order derivative matrix}: 
\begin{align*}\label{eq:WC_Beam_Eij}
\bar{D}_{ij}=\sum_{j=1}^{N+6}\sum_{k=1}^{N} B_{ik}\bar{B}_{kj}  \,\,\,\,\,\,\, (i=1,2,...,N)\tag{23}
\end{align*} 

\noindent \textit{Fifth order derivative matrix}: 

\begin{align*} \label{eq:WC_Beam_D1ij}
{V}_{ij}=\begin{cases}
\bar{D}_{ij} \,\,\,\, (i=2,3,\cdots,N-1; \,\, j=1,2,\cdots,N)\\ \\
0 \,\,\,\,\,\,\,(i=2,3,\cdots,N-1; \,\, j=N+1,\cdots,N+6)\tag{24}
\end{cases} 
\end{align*}
\begin{align*} \label{eq:WC_Beam_D2ij}
{V}_{ij}=\sum_{k=2}^{N-1}B_{ik}B_{kj} \,\,\, (i=1,N; \,\,\,\, j=1,2,\cdots,N)\\ 
{V}_{i(N+3)}=B_{i1}\,\,; \,\,\, \, V_{i(N+4)}=B_{iN} \,\,\,(i=1,N)\tag{25}
\end{align*} 
\begin{align*}\label{eq:WC_Beam_Eij}
\bar{E}_{ij}=\sum_{k=1}^{N} A_{ik}{V}_{kj}  \,\,\,\,\,(i=1,2,...,N; \,\,j=1,2,...,N+6)\tag{26}
\end{align*} 

\noindent \textit{Sixth order derivative matrix}: 
\begin{align*} \label{eq:WC_Beam_Fij}
\bar{F}_{ij}=\sum_{k=1}^{N} B_{ik}{V}_{kj}  \,\,\,\,\,(i=1,2,...,N; \,\,j=1,2,...,N+6)\tag{27}
\end{align*}

\noindent \textit{Seventh order derivative matrix}: 
\begin{align*}\label{eq:WC_Beam_Eij}
\tilde{E}_{ij}=\sum_{j=1}^{N+6}\sum_{k=1}^{N} C_{ik}\bar{C}_{kj}  \,\,\,\,\,\,\, (i=1,2,...,N)\tag{28}
\end{align*} 

\begin{align*} \label{eq:WC_Beam_D1ij}
{Y}_{ij}=\begin{cases}
\tilde{E}_{ij} \,\,\,\, (i=2,3,\cdots,N-1; \,\, j=1,2,\cdots,N)\\ \\
0 \,\,\,\,\,\,\,(i=2,3,\cdots,N-1; \,\, j=N+1,\cdots,N+6)\tag{29}
\end{cases} 
\end{align*}
\begin{align*} \label{eq:WC_Beam_D2ij}
{Y}_{ij}=\sum_{k=2}^{N-1}C_{ik}C_{kj} \,\,\, (i=1,N; \,\,\,\, j=1,2,\cdots,N)\\ 
{Y}_{i(N+5)}=C_{i1}\,\,; \,\,\, \, Y_{i(N+6)}=C_{iN} \,\,\,(i=1,N)\tag{30}
\end{align*} 
\begin{align*}\label{eq:WC_Beam_Eij}
\bar{G}_{ij}=\sum_{k=1}^{N} A_{ik}{Y}_{kj}  \,\,\,\,\,(i=1,2,...,N; \,\,j=1,2,...,N+6)\tag{31}
\end{align*} 

\noindent \textit{Eight order derivative matrix}: 
\begin{align*} \label{eq:WC_Beam_Fij}
\bar{H}_{ij}=\sum_{k=1}^{N} B_{ik}{Y}_{kj}  \,\,\,\,\,(i=1,2,...,N; \,\,j=1,2,...,N+6)\tag{32}
\end{align*}

Here, $\bar{A}_{ij}$, $\bar{B}_{ij}$, $\bar{C}_{ij}$, $\bar{D}_{ij}$, $\bar{E}_{ij}$, $\bar{F}_{ij}$,  $\bar{G}_{ij}$ and  $\bar{H}_{ij}$ are first to eight order modified weighting coefficients matrices, respectively. Using the above Equations (\ref{eq:WC_Beam_Aij})-(\ref{eq:WC_Beam_Fij}), the governing differential Equation (\ref{EOM_beam}), at inner grid points interms of the differential quadrature framework is written as
 \begin{align*}         
	 EI\sum_{j=1}^{N+6}\bar{D}_{ij}\bar{w}_{j}-g_{1}^{2}EI\sum_{j=1}^{N+6}\bar{F}_{ij}\bar{w}_{j}+
	 g_{2}^{4}EI\sum_{j=1}^{N+6}\bar{H}_{ij}\bar{w}_{j}=
	  q(x_{i})+P\bar{A}_{ij}+m_{i}\ddot{w} \\ \,\,\,\,\,\, (i=2,3,...,N-1)\tag{33}
\end{align*}

The boundary forces given by Equations (\ref{eq:BC_Cl_Beam})-(\ref{eq:BC_NCl_Beam}), are expressed as \\

\noindent Shear force:
 \begin{align*} \label{eq:Boundary_forces_Beam_V}        
	 V_{i}=EI\sum_{j=1}^{N+6}\bar{C}_{ij}\bar{w}_{j}-g_{1}^{2}EI\sum_{j=1}^{N+6}\bar{E}_{ij}\bar{w}_{j}+
	 g_{2}^{4}EI\sum_{j=1}^{N+6}\bar{G}_{ij}\bar{w}_{j} \,\,\,\,\,\, (i=1,N)\tag{34}
\end{align*} 
\noindent Bending moment:
 \begin{align*} \label{eq:Boundary_forces_Beam_M}        
	 M_{i}=EI\sum_{j=1}^{N+6}\bar{B}_{ij}\bar{w}_{j}-g_{1}^{2}EI\sum_{j=1}^{N+6}\bar{D}_{ij}\bar{w}_{j}+
	 g_{2}^{4}EI\sum_{j=1}^{N+6}\bar{E}_{ij}\bar{w}_{j} \,\,\,\,\,\, (i=1,N)\tag{35}
\end{align*} 
\noindent Double moment:
 \begin{align*} \label{eq:Boundary_forces_Beam_MM}        
	 \bar{M}_{i}=g_{1}^{2}EI\sum_{j=1}^{N+6}\bar{C}_{ij}\bar{w}_{j}-g_{2}^{4}EI\sum_{j=1}^{N+6}\bar{E}_{ij}\bar{w}_{j} \,\,\,\,\,\ (i=1,N)
\tag{36}
\end{align*}  
\noindent Triple moment:
 \begin{align*} \label{eq:Boundary_forces_Beam_MMM}        
\bar{\bar{M}}_{i}=g_{2}^{4}EI\sum_{j=1}^{N+6}\bar{D}_{ij}\bar{w}_{j} \,\,\,\,\,\ (i=1,N)\tag{37}
\end{align*}  

\noindent here $i=1$ and $i=N$ correspond to the left support $x=0$ and right support $x=L$ of the beam, respectively.

After applying the respective boundary conditions the system of equations are given as:

\begin{align*}\label{eq:Boundary_disp_Beam}
\begin{bmatrix}
k_{bb} & \phantom{-}k_{bd} \\ \\ 
k_{db} & \phantom{-}k_{dd} \\ \\ 
 \end{bmatrix}\begin{Bmatrix}
\Delta_{b} \\ \\ 
\Delta_{d}  \\ \\ 
 \end{Bmatrix}=\begin{Bmatrix}
f_{b} \\ \\ 
f_{d} 
 \end{Bmatrix}+\begin{bmatrix}
I & \phantom{-}0 \\ \\ 
\,0 & \phantom{-}\omega^{2}M_{dd} \\ \\ 
 \end{bmatrix}\begin{Bmatrix}
0 \\ \\ 
\Delta_{d} 
 \end{Bmatrix} +P\begin{bmatrix}
0 & \phantom{-}0 \\ \\ 
G_{db} & \phantom{-}G_{dd} \\ \\ 
 \end{bmatrix}\begin{Bmatrix}
\Delta_{b} \\ \\ 
\Delta_{d} 
 \end{Bmatrix} \tag{38}
\end{align*}

\noindent where the subscript $b$ and $d$ indicates the boundary and domain of the beam element. $f_{b}$, $\Delta_{b}$ and $f_{d}$, $\Delta_{d}$ are the boundary and domain forces and displacements of the beam, respectively. For static analysis the system of equations are written as 

\begin{align*}\label{eq:Boundary_disp_Beam}
\begin{bmatrix}
\,k_{bb} & \phantom{-}k_{bd} \\ \\ 
\,k_{db} & \phantom{-}k_{dd} \\ \\ 
 \end{bmatrix}\begin{Bmatrix}
\,\Delta_{b} \\ \\ 
\,\Delta_{d}  \\ \\ 
 \end{Bmatrix}=\begin{Bmatrix}
\,f_{b} \\ \\ 
\,f_{d}  \\ \\ 
 \end{Bmatrix} \tag{39}
\end{align*}

Expressing the system of equations in terms of domain dofs $\Delta_{d}$, we obtain
\begin{align*} \label{eq:Domain_disp_Beam}
\Big[k_{dd}-k_{db}k_{bb}^{-1}k_{bd}\Big] {\Big\{}\Delta_{d}{\Big\}}=
{\Big\{}f_{d}-k_{db}k_{bb}^{-1}f_{b}{\Big\}}\tag{40}
\end{align*}

The solution of the above system of equations gives the unknown displacements at the domain nodes of the beam element. The boundary displacements are computed from Equation (\ref{eq:Boundary_disp_Beam}), and forces are computed from the Equations (\ref{eq:Boundary_forces_Beam_V})-(\ref{eq:Boundary_forces_Beam_MM}).

\noindent Similarly for free vibration analysis $f_{b}=f_{d}=P=0$, and the system of equations are reduced to

\begin{align*}\label{eq:free_vib_eq}
\begin{bmatrix}
k_{bb} & \phantom{-}k_{bd} \\ \\ 
k_{db} & \phantom{-}k_{dd} \\ \\ 
 \end{bmatrix}\begin{Bmatrix}
\Delta_{b} \\ \\ 
\Delta_{d}  \\ \\ 
 \end{Bmatrix}=\begin{bmatrix}
I & \phantom{-}0 \\ \\ 
\,0 & \phantom{-}\omega^{2}M_{dd} \\ \\ 
 \end{bmatrix}\begin{Bmatrix}
0 \\ \\ 
\Delta_{d} 
 \end{Bmatrix} \tag{41}
\end{align*}

and finally for stability analysis the system of equations are given by

\begin{align*}\label{eq:Buck_eq}
\begin{bmatrix}
k_{bb} & \phantom{-}k_{bd} \\ \\ 
k_{db} & \phantom{-}k_{dd} \\ \\ 
 \end{bmatrix}\begin{Bmatrix}
\Delta_{b} \\ \\ 
\Delta_{d}  \\ \\ 
 \end{Bmatrix}=P\begin{bmatrix}
0 & \phantom{-}0 \\ \\ 
G_{db} & \phantom{-}G_{dd} \\ \\ 
 \end{bmatrix}\begin{Bmatrix}
\Delta_{b} \\ \\ 
\Delta_{d} 
 \end{Bmatrix} \tag{42}
\end{align*}

The Equations (\ref{eq:free_vib_eq}) and (\ref{eq:Buck_eq}) represents an Eigen value problem and the solution provides the frequencies and buckling load.

\section{Numerical Results and Discussion}
The accuracy and convergence characteristic of the proposed differential quadrature beam element is verified through numerical examples on static, free vibration and stability analysis. The results are compared with the analytical solutions obtained in the Appendix-I for two different combinations of length scale parameters, $g_{1}=0.1, g_{2}=0.05$ and $g_{1}=0.15, g_{2}=0.1$. Single DQ element is used in the present study. The grid employed in the present analysis is unequal Gauss\textendash Lobatto\textendash Chebyshev points given by
\begin{align*}    
x_{i}=\frac{1}{2}\Bigg[1-cos{\frac{(i-1)\pi}{N-1}} \Bigg] \tag{43}
\end{align*}

\noindent where $N$ is the number of grid points and $x_{i}$ are the coordinates of the grid. 

The  classical and non-classical boundary conditions used in this study for different end supports are listed in the Section \ref{Sg_formulation}. The non-classical boundary conditions employed for simply supported gradient beam are $w^{''}=w^{'''}=0$ at $x=(0,L)$, the equations related to curvature and triple displacement derivative are eliminated. For the cantilever beam the non-classical boundary conditions used are $w^{''}=w^{'''}=0$ at $x=0$ and $\bar{M}=\bar{\bar{M}}=0$ at $x=L$. The equation related to curvature and triple displacement derivative at $x=0$ are eliminated and the equation related to double and triple  moment at $x=L$ are retained. Similarly, for clamped and propped cantilever beam the non-classical boundary conditions remains the same, $w^{''}=w^{'''}=0$ at $x=(0,L)$. The numerical data used for the analysis of beams is as follows: Length $L=1$, Young's modulus $E=3 \times 10^{6}$, Poission's ratio $\nu=0.3$, density $\rho =1$ and load $q=1$. 

\subsection{Static analysis of second strain gradient Euler-Bernoulli beam}

In this section, the capability of the element is demonstrated for static analysis of gradient elastic beams subjected to uniformly distributed load. Three support conditions are considered in this analysis, simply supported, clamped and cantilever. The performance of the element is verified by comparing the classical (deflection and slope) and the non-classical (curvature, triple displacement derivative, double moment and triple moment) quantities with the exact values. The results reported here for beams with udl are nondimensional as,  deflection : $\bar{w}=100EIw/qL^{4}$, bending moment: $B_{m}=M/qL^{2}$, curvature :$\bar{w}^{''}=w^{''}L$, triple derivative of displacement :$\bar{w}^{'''}=w^{'''}L^{2}$, double moment : $D_{m}=\bar{M}/qL^{3}$ and triple moment : $T_{m}=\bar{\bar{M}}/qL^{4}$. 

\setlength{\extrarowheight}{.5em}
\begin{table}[H]
    \centering
    \caption{Comparison of deflection, slope, curvature, double and triple  moment for a simply supported beam under a udl.}
      
 \begin{tabular}{c@{\hskip 0.01in}c@{\hskip 0.1in}c@{\hskip 0.1in}c@{\hskip 0.1in}c@{\hskip 0.1in}c@{\hskip 0.1in}c@{\hskip 0.1in}}
     \\ \hline
 & $ \text{N}$     & ${w}_{\,(x=L/2)}$ 	         & $w^{'}_{\,(x=0)}$      & ${w}^{''}_{\,(x=L/2)}$  & $D_{m\,(x=0)}$ & $T_{m\,(x=0)}$ \\
& &    & $\times 10^{-3}$      & $\times 10^{-3}$       & $\times 10^{-3}$  & $\times 10^{-3}$ \\ \hline 
   \multirow{3}{*}{} &  \multicolumn{1}{c}{}  & \multicolumn{1}{c}{} &  \multicolumn{1}{c}{$(\frac{g_{1}}{L}=0.1$,}& \multicolumn{1}{c}{$\frac{g_{2}}{L}=0.05)$}& \multicolumn{1}{c}{}& \multicolumn{1}{c}{} \\    
\cline{4-5}

&$5$ 	          &  1.1242  & 0.1397  & 0.4401  & 3.6074 & 0.0063 \\

&$7$ 	           &  1.1656  & 0.1471  & 0.4540   & 3.6884 & 0.0061  \\ 

&$9$ 	         &  1.1631  & 0.1463  & 0.4537 & 3.4360 & 0.0113 \\ 

 &$11$           &  1.1479  & 0.1432  & 0.4501 & 3.6968 & 0.0599  \\

& $13$  	          &  1.1559  & 0.1438  & 0.4537  & 4.5624 & 0.1118   \\

&$15$  	          &  1.1676  & 0.1455  & 0.4578  & 4.9474  & 0.1245  \\ 
&$17$  	          &  1.1724  & 0.1462  & 0.4593  & 5.0214 & 0.1245  \\ 
&$19$  	          &  1.1739  & 0.1465  & 0.4597  & 5.0277 & 0.1226 \\ 
&$21$  	          &  1.1742  & 0.1465  & 0.4598  & 5.0263 & 0.1220  \\ 

& $\text{Exact}$   & 1.1743      &  0.1465 & 0.4598 & 5.0252 & 0.1218 	 \\ \hline 

   \multirow{3}{*}{} &  \multicolumn{1}{c}{}  & \multicolumn{1}{c}{} &  \multicolumn{1}{c}{$(\frac{g_{1}}{L}=0.15,$}& \multicolumn{1}{c}{$\frac{g_{2}}{L}=0.1)$}& \multicolumn{1}{c}{}& \multicolumn{1}{c}{} \\    
\cline{4-5}   

&$5$ 	     &  0.9463  & 0.1154  & 0.3749  & 6.7135 & 0.0015  \\

&$7$ 	          &  0.9562  & 0.1164  & 0.3854  & 7.9528 & 0.2522 \\ 

&$9$ 	         &  0.9614  & 0.1155  & 0.3934 & 11.1079 & 0.7041   \\ 

 &$11$          &  0.9883  & 0.1194  & 0.4025  & 11.6260 & 0.7199  \\

& $13$  	          &  0.9931  & 0.1201  & 0.4039  & 11.6126 & 0.7056 \\

&$15$  	         &  0.9936  & 0.1202  & 0.4040  & 11.6054 & 0.7054  \\ 
&$17$  	          &  0.9936  & 0.1202  & 0.4040 & 11.6045 & 0.7052 \\ 
&$19$  	          &  0.9936  & 0.1202  & 0.4040  & 11.6045 & 0.7052 \\
&$21$  	          &  0.9936  & 0.1202  & 0.4040  & 11.6045 & 0.7052 \\ 
&$\text{Exact}$   & 0.9936      &  0.1202 & 0.4040 & 11.6045 & 0.7052 \\ \hline 

        \end{tabular}
    \label{Delf_SS_udl}
\end{table}

\setlength{\extrarowheight}{.5em}
\begin{table}[H]
    \centering
    \caption{Comparison of deflection, curvature, double and triple moment for a clamped beam under a udl.}
      
 \begin{tabular}{c@{\hskip 0.01in}c@{\hskip 0.1in}c@{\hskip 0.1in}c@{\hskip 0.1in}c@{\hskip 0.1in}c@{\hskip 0.1in}}
     \\ \hline
 & $ \text{N}$       & ${w}_{\,(x=L/2)}$ & ${w}^{''}_{\,(x=L/2)}$  & $D_{m\,(x=0)}$ & $T_{m\,(x=0)}$ \\
& &    & $\times 10^{-3}$        & $\times 10^{-3}$  & $\times 10^{-3}$ \\ \hline 
   \multirow{3}{*}{} &  \multicolumn{1}{c}{}  & \multicolumn{1}{c}{} &  \multicolumn{1}{c}{$(\frac{g_{1}}{L}=0.1,$}& \multicolumn{1}{c}{$\frac{g_{2}}{L}=0.05)$}& \\    
\cline{4-5}

&$5$ 	          &  0.1133  & 0.0994  & 4.2731 & 0.0137 \\

&$7$ 	           &  0.0987  & 0.0900   & 5.1124 & 0.0518  \\ 

&$9$ 	         &  0.0967  & 0.0865 & 4.5254 & 0.0483 \\ 

 &$11$           &  0.0854  & 0.0800 & 5.1261 & 0.0278 \\

& $13$  	          &  0.0784 & 0.0761  & 6.4657 & 0.1115   \\

&$15$  	          &  0.0789 & 0.0764 & 6.8750  & 0.1278  \\ 
&$17$  	          &  0.0802  & 0.0771  & 6.8948 & 0.1248  \\ 
&$19$  	          &  0.0808  & 0.0774  & 6.8743 & 0.1222 \\ 
&$21$  	          &  0.0810    & 0.0776  & 6.8642 & 0.1211  \\ 

& $\text{Exact}$   & 0.0810       & 0.0776 & 6.8604 & 0.1208	 \\ \hline 

   \multirow{3}{*}{} &  \multicolumn{1}{c}{}  & \multicolumn{1}{c}{} &  \multicolumn{1}{c}{$(\frac{g_{1}}{L}=0.15,$}& \multicolumn{1}{c}{$\frac{g_{2}}{L}=0.1)$}& \multicolumn{1}{c}{} \\    
\cline{4-5}   

&$5$ 	     &  0.0390   & 0.3749  & 4.6644 & 0.2576  \\

&$7$ 	          &  0.0520    & 0.0520  & 4.9381 & 0.3009\\ 

&$9$ 	         &  0.0268  & 0.0320 & 9.6015 & 0.3329   \\ 

 &$11$          &  0.0295  & 0.0340  & 9.6332 & 0.3415  \\

& $13$  	          &  0.0311   & 0.0352  & 9.4460 & 0.3174 \\

&$15$  	         &  0.0313   & 0.0354  & 9.4153 & 0.3133  \\ 
&$17$  	          &  0.0313   & 0.0354 & 9.4124 & 0.3129\\ 
&$19$  	          &  0.0313  & 0.0354  & 9.4123 & 0.3129 \\
&$21$  	          &  0.0313  & 0.0354   & 9.4123 & 0.3129 \\ 

&$\text{Exact}$   & 0.0313    & 0.0354 & 9.4123 & 0.3129 \\ \hline 

        \end{tabular}
    \label{Delf_clamped_udl}
\end{table}

\setlength{\extrarowheight}{.5em}
\begin{table}[H]
    \centering
    \caption{Comparison of deflection, slope, curvature, triple displacement derivative, double and triple moment for a cantilever beam under a udl.}
      
 \begin{tabular}{c@{\hskip 0.01in}c@{\hskip 0.1in}c@{\hskip 0.1in}c@{\hskip 0.1in}c@{\hskip 0.1in}c@{\hskip 0.1in}c@{\hskip 0.1in}c@{\hskip 0.1in}}
     \\ \hline
 & $ \text{N}$     & ${w}_{\,(x=L)}$ 	         & $w^{'}_{\,(x=L)}$      & ${w}^{''}_{\,(x=L/2)}$ & ${w}^{'''}_{\,(x=L/2)}$ & $D_{m\,(x=0)}$ & $T_{m\,(x=0)}$ \\
& &    & $\times 10^{-3}$      & $\times 10^{-3}$       & $\times 10^{-3}$  & $\times 10^{-3}$& $\times 10^{-3}$ \\ \hline 
   \multirow{3}{*}{} &  \multicolumn{1}{c}{}  & \multicolumn{1}{c}{} &  \multicolumn{1}{c}{$(\frac{g_{1}}{L}=0.1$,}& \multicolumn{1}{c}{$\frac{g_{2}}{L}=0.05)$}& \multicolumn{1}{c}{}& \multicolumn{1}{c}{} \\    
\cline{4-5}

&$5$ 	          &  7.3448  & 0.3876  & 0.3636  & 2.4728 & 30.9229 & 0.0668 \\

&$7$ 	           &  7.5689  & 0.4380  & 0.4491   & 1.7241 & 37.1076 & 0.2550  \\ 

&$9$ 	         &  8.1486  & 0.5067  & 0.5299 & 1.7889 & 33.0394 & 0.2136 \\ 

 &$11$           &  7.8110  & 0.4891  & 0.5399 & 1.9633 & 37.2511 & 0.3858  \\

& $13$  	          &  7.5423  & 0.4698  & 0.5330 & 1.5194  & 47.0093 & 0.9621   \\

&$15$  	          &  7.5465  & 0.4658  & 0.5279 & 1.8599 & 49.9742  & 1.0436 \\ 
&$17$  	          &  7.5837  & 0.4661  & 0.5259  & 1.8391 & 50.1746 & 1.0103  \\ 
&$19$  	          &  7.6023  & 0.4666  & 0.5253  & 1.8458 & 50.0539 & 0.9870 \\ 
&$21$  	          &  7.6088  & 0.4668  & 0.5251  & 1.8450 & 49.9893 & 0.9782 \\ 

& $\text{Exact}$   & 7.6106      &  0.4668 & 0.5251 & 1.8453 & 49.9620 & 0.9750 	 \\ \hline 

   \multirow{3}{*}{} &  \multicolumn{1}{c}{}  & \multicolumn{1}{c}{} &  \multicolumn{1}{c}{$(\frac{g_{1}}{L}=0.15,$}& \multicolumn{1}{c}{$\frac{g_{2}}{L}=0.1)$}& \multicolumn{1}{c}{}& \multicolumn{1}{c}{} \\    
\cline{4-5}   

&$5$ 	     &  4.9134  & 0.2996  & 0.3520  & 1.9194 & 40.1363 & 1.1514  \\

&$7$ 	          &  6.0607  & 0.4268  & 0.5255 & 1.5420 & 48.8664 & 0.6205 \\ 

&$9$ 	         &  5.0586  & 0.3674 & 0.5204 & 1.1617 & 78.6422 & 4.0284   \\ 

 &$11$          &  5.2501  & 0.3633  & 0.5048  & 1.3175 & 76.8798 & 3.5652 \\

& $13$  	          &  5.3320  & 0.3653  & 0.5027  & 1.3299 & 75.4694& 3.3091 \\

&$15$  	         &  5.3453  & 0.3657  & 0.5025  & 1.3331 & 75.2408 & 3.2678  \\ 
&$17$  	          &  5.3460  & 0.3657  & 0.5025 & 1.3333 & 75.2120 & 3.2635 \\ 
&$19$  	          &  5.3454  & 0.3657  & 0.5024  & 1.3332 & 75.2040 & 3.2629 \\
&$21$  	          &  5.3596  & 0.3668  & 0.5039  & 1.3352 & 75.3480 & 3.2695 \\ 
&$\text{Exact}$   & 5.3437     &  0.3658 & 0.5022 & 1.3357 & 75.2160 & 3.2634 \\ \hline 

        \end{tabular}
    \label{Delf_cantilever_udl}
\end{table}

In Table \ref{Delf_SS_udl}, convergence of nondimensional deflection, slope, curvature, double and triple moment are given for two different combinations of length scale values $g_{1}/L=0.1$, $g_{2}/L=0.05$  and $g_{1}/L=0.15$, $g_{2}/L=0.1$. The results are compared with exact solutions obtained in Appendix-I. The deflection and curvature are computed at the center of the beam $x=L/2$, the slope, double moment and triple moment at $x=0$. The convergence is seen faster for both classical and non-classical quantities obtained using the present element. Similar convergence trend is noticed in Tables \ref{Delf_clamped_udl}-\ref{Delf_cantilever_udl} for clamped and cantilever beam. A good agreement with the exact solution is seen with 15 grid points for $g_{1}/L=0.1$, $g_{2}/L=0.05$ values and converged solutions are obtained using 21 grid points. For $g_{1}/L=0.15$, $g_{2}/L=0.1$ converged solution are obtained using 15 grid points.

\setlength{\extrarowheight}{.3em}
\begin{table}[H]
    \centering
    \caption{Comparison of deflection, slope, curvature and triple displacement derivative for a simply supported beam along the length.}
      
 \begin{tabular}{c@{\hskip 0.15in}c@{\hskip 0.15in}c@{\hskip 0.15in}c@{\hskip 0.15in}c@{\hskip 0.15in}c@{\hskip 0.15in}c@{\hskip 0.15in}c@{\hskip 0.15in}c@{\hskip 0.1in}}
     \\ \hline
 &$w$&  & $w^{'}$ &  &$w^{''}$ & &$w^{'''}$   \\
$x/L $ & present& Exact &present  & Exact  & present  & Exact  & present & Exact  \\ \hline 
 
 0.0000&  0.0000  & 0.0000  & 0.1202  & 0.1507 &  0.0000  & 0.0000  & 0.0000 & 0.0000   \\

0.0125&  0.0377  & 0.0377  & 0.1202  & 0.1502  &  -0.0021  & -0.0021  & -0.3187 & -0.3191 \\

0.0495&  0.1487  & 0.1487  & 0.1198  & 0.1435  &  -0.0266  & -0.0266  & -0.9373 & -0.9375 \\

0.1091&  0.3249  & 0.3249 & 0.1162  & 0.1188 &  -0.0961  & -0.0961  & -1.3049 & -1.3050 \\

0.1882&     0.5447  & 0.5447  & 0.1045  & 0.0687  &  -0.1997  & -0.1997  & -1.2507 & -1.2506  \\

0.2830  & 0.7657  & 0.7657  & 0.0804 &  0.0000  & -0.3036  & -0.3036   & -0.9197 & -0.9196\\

0.3887 & 0.9317  & 0.9317 &  0.0440  & -0.0687  & -0.3775  & -0.3775 & -0.4758 & -0.4758\\

0.5000  &0.9936 & 0.9936  & 0.000 & -0.1188   & -0.4040  & -0.4040  & 0.0000 & 0.0000\\

0.6113  &0.9317 & 0.9317  &-0.0440  & -0.1435   & -0.3776  & -0.3777  &0.4758 & 0.4758 \\

0.7169  &0.7657 & 0.7657  & -0.0804 &  -0.1502  & -0.3036  & -0.3036  & 0.9197 & 0.9196 \\

0.8117  &0.5447 & 0.5447  & -0.1045 & -0.1507  & -0.1997  & -0.1997  & 1.2507 & 1.2506 \\

0.8909  &0.3249 & 0.3249  & -0.1162 & -0.1507  & -0.0961  & -0.0961  & 1.3049 & 1.3049 \\

0.9505  &0.1487 & 0.1487  & -0.1198 & -0.1507  & -0.0266  & -0.0266  & 0.9373 & 0.9375 \\

0.9875  &0.0377 & 0.0377  & -0.1202 & -0.1507  & -0.0021  & -0.0021  & 0.3187 & 0.3191 \\

1.0000  &0.0000 & 0.0000  & -0.1202 & -0.1507  & 0.0000  & 0.0000  & 0.0000 & 0.0000 \\

 \hline 

        \end{tabular}
    \label{Defl_along_length_SS2_udl}
\end{table}

In Table \ref{Defl_along_length_SS2_udl}, comparison is made for classical and non-classical quantities computed along the length of a simply supported beam subjected to udl. The results are obtained using 15 grid points for $g_{1}/L=0.15$, $g_{2}/L=0.1$ values. Excellent match with the exact solutions is exhibited for all the classical and non-classical quantities along the length of the beam.

From the above tabulated results it can be concluded that the solutions obtained using the proposed element with 15 grid points are in excellent agreement with the exact solutions for all the boundary conditions and $g/L$ values considered. Hence, a single element with fewer nodes can be efficiently applied to study the static behaviour of a second strain gradient Euler-Bernoulli beam for any choice of intrinsic length and boundary condition.
\subsection{Free vibration analysis of gradient elastic beams}

The applicability of the  proposed beam element for free vibration analysis of second strain gradient beam will be verified in this section. The first six elastic frequencies obtained for different boundary conditions are compared with the analytical solutions computed in the Appendix-I for $g_{1}/L=0.1$, $g_{2}/L=0.05$ and $g_{1}/L=0.15$, $g_{2}/L=0.1$. Four different boundary conditions are considered in this analysis, simply supported, clamped, cantilever and free-free. In Table \ref{Freq_ss_II_beam}, convergence behaviour of the first six frequencies for a simply supported gradient beam are shown. The frequencies obtained using 15 grid point are in close agreement with the analytical solutions for both combinations of $g/L$ values. Similar convergence trend and accuracy is noticed in the Tables \ref{Freq_clamped_beam}-\ref{Freq_Free_Free_beam}, for clamped, cantilever and free-free beams, respectively. 

\setlength{\extrarowheight}{.5em}
\begin{table}[H]
    \centering
    \caption{Comparison of first six frequencies for a simply supported gradient beam.}
      
 \begin{tabular}{c@{\hskip 0.01in}c@{\hskip 0.1in}c@{\hskip 0.1in}c@{\hskip 0.1in}c@{\hskip 0.1in}c@{\hskip 0.1in}c@{\hskip 0.1in}c@{\hskip 0.1in}}
     \\ \hline
 & N & $\bar{\omega}_{1}$    & $\bar{\omega}_{2}$  & $\bar{\omega}_{3}$& $\bar{\omega}_{4}$ &$\bar{\omega}_{5}$ &$\bar{\omega}_{6}$
\\ \hline 

   \multirow{3}{*}{} &  \multicolumn{1}{c}{}  & \multicolumn{1}{c}{} &  \multicolumn{1}{c}{$(\frac{g_{1}}{L}=0.1$,}& \multicolumn{1}{c}{$\frac{g_{2}}{L}=0.05)$}& \multicolumn{1}{c}{}& \multicolumn{1}{c}{} \\    
\cline{4-5}
  
&$5$&10.2230     & 74.7250  & 367.1055  & --- & ---  & ---   \\

&$7$& 10.3637  &  45.6637  &  102.3530  & 525.1573  & 680.9263 &--- \\

&$9$&10.4383 &  47.6502 & 127.1765  & 267.7521&  357.8939 & 2311.9832   \\ 

 &$11$ & 10.515         & 48.3026  & 127.6144  & 266.3728  & 474.5031 & 1364.4916  \\

&$13$& 10.4838  	 & 48.1502 &  128.1376  & 277.3628  & 526.9411 & 776.2329   \\

&$15$  & 10.4340	          & 47.8289  &  127.7450  & 272.6296  &  505.0123 & 920.0973   \\ 
&$17$  	 &  10.4134         &  47.6803  &  127.4444  &  271.7701  & 506.5422 & 854.8488 \\ 
&$19$  	& 10.4073          &  47.6300  & 127.3290  & 271.3294  & 505.6394 &  862.4963   \\
&$21$  	&10.4058          & 47.6156  &  127.2946  & 271.2002  & 505.4833 &860.6482   \\  
& Analyt.   &10.4058          & 47.6156  &  127.2946  & 271.1597  & 505.4257 &860.6195  	 \\ \hline 

   \multirow{3}{*}{} &  \multicolumn{1}{c}{}  & \multicolumn{1}{c}{} &  \multicolumn{1}{c}{$(\frac{g_{1}}{L}=0.15,$}& \multicolumn{1}{c}{$\frac{g_{2}}{L}=0.1)$}& \multicolumn{1}{c}{}& \multicolumn{1}{c}{} \\    
\cline{4-5}   

&$5$&10.7250  & 126.6505  & 138.7740  & --- & ---  & ---   \\

&$7$&  11.4542 & 78.7448  &  160.1790  &  438.9373  & 769.3881 &--- \\ 

&$9$&11.5201 & 63.8629 & 183.8644  & 643.4299 &  1473.2662 &  2918.2181  \\ 

& $11$&11.3653 	 & 62.2388 &  197.0683  & 424.9276  &  736.9888 & 3072.7803   \\
&$13$  	& 11.3372  & 61.6128  & 193.7879 & 489.2711  & 1172.3459 &  1448.7316 \\ 
&$15$  	 & 11.3342        & 61.5469  & 193.8910 & 472.0980 & 970.9535 & 2234.6097 \\ 
&$17$  	 & 11.3340   &   61.5405 &  193.8719  & 473.9725 & 991.7657 &  1801.7299  \\
&$19$  	 & 11.3340   &  61.5401  & 193.8715  & 473.8101 & 989.1708 &  1860.3001  \\ 
&$21$  	 & 11.3340   &  61.5401  &  193.8714  &  473.8181 & 989.3820 &  1850.6519  \\ 

& Analyt. & 11.3340   &  61.5401  &  193.8714  &  473.8175 & 989.3680 &  1851.5906	 \\ \hline

        \end{tabular}
    \label{Freq_ss_II_beam}
\end{table}

\setlength{\extrarowheight}{.5em}
\begin{table}[H]
    \centering
    \caption{Comparison of first six frequencies for a clamped gradient beam.}
      
 \begin{tabular}{c@{\hskip 0.01in}c@{\hskip 0.1in}c@{\hskip 0.1in}c@{\hskip 0.1in}c@{\hskip 0.1in}c@{\hskip 0.1in}c@{\hskip 0.1in}c@{\hskip 0.1in}}
     \\ \hline
 & N & $\bar{\omega}_{1}$    & $\bar{\omega}_{2}$  & $\bar{\omega}_{3}$& $\bar{\omega}_{4}$ &$\bar{\omega}_{5}$ &$\bar{\omega}_{6}$
\\ \hline 

   \multirow{3}{*}{} &  \multicolumn{1}{c}{}  & \multicolumn{1}{c}{} &  \multicolumn{1}{c}{$(\frac{g_{1}}{L}=0.1$,}& \multicolumn{1}{c}{$\frac{g_{2}}{L}=0.05)$}& \multicolumn{1}{c}{}& \multicolumn{1}{c}{} \\    
\cline{4-5}
  
&$5$& 49.9249    &  49.9249 &  75.8825  & --- & ---  & ---   \\

&$7$&  36.1724  &  149.5939  &  256.7750  &   256.7750  & 417.2657 &--- \\

&$9$&  37.1702 &   117.8567 &  231.6243  &  1054.0463 &1120.8910 & 1120.8910 \\ 

 &$11$& 39.4632 &   125.4928 &  262.9499  &  759.4087 &876.5466 & 2153.0935 \\

&$13$&  41.1352 &   131.1305 &  288.4611  &  511.1651 &745.5959 & 2792.7604   \\

&$15$  &  40.9844         & 127.7012  &  283.4289 & 565.7662  & 1184.2651 &  1184.2651   \\ 
&$17$  	 &  40.6729         &  125.6902  &   280.7206  &  535.0132  & 904.316 &  1799.6178 \\ 
&$19$  	& 40.5314          & 124.9045 &  279.3736  &  532.1407  & 914.4509 &   143.1897   \\
&$21$  	& 40.4857        &124.6561 & 278.9482  & 530.5742  & 910.6078 & 1462.6078   \\  
&Analyt.& 40.4857       &124.6561 & 278.9482  & 530.1349  & 910.2262 & 1455.7444 	 \\ \hline 
   \multirow{3}{*}{} &  \multicolumn{1}{c}{}  & \multicolumn{1}{c}{} &  \multicolumn{1}{c}{$(\frac{g_{1}}{L}=0.15,$}& \multicolumn{1}{c}{$\frac{g_{2}}{L}=0.1)$}& \multicolumn{1}{c}{}& \multicolumn{1}{c}{} \\    
\cline{4-5}   
&$5$&53.6137  & 53.6137   & 88.3409  & --- & ---  & ---   \\

&$7$&  51.6586 &  150.3018  &   377.3499  & 377.3499  & 807.8832 &--- \\ 

&$9$& 69.1149 &  242.0681 & 898.0838 &  898.0838 & 1055.0828 & 3200.8126 \\ 

& $11$& 67.3022 	 &  231.4358 & 482.3860  & 1251.1029  &  2401.0403 &  8404.3042   \\
&$13$  	&65.6599 & 226.28010 & 575.4737 &  958.2332 &  1370.1105 &  5498.4399 \\ 
&$15$  	 &  65.4442        & 221.6120 & 542.5355 & 1223.8458 &  2498.9469 & 2493.0270 \\ 
&$17$  	 & 65.4249  &   221.4972 & 545.1122  & 1117.9998 & 2006.7471 &   4623.3352  \\
&$19$  	 & 65.4237   &  221.4775	  & 544.9018 & 1130.6769 & 2106.0718 &  3404.5865  \\ 
&$21$  	 & 65.4236   & 221.4765	  & 544.9096 & 1129.3365 & 2090.5403 &  3609.1100  \\ 

&Analyt.  & 65.4235   & 221.4764	  & 544.9086 & 1129.4247 & 2091.9224 &  3573.1262 	 \\ \hline

        \end{tabular}
    \label{Freq_clamped_beam}
\end{table}

\setlength{\extrarowheight}{.5em}
\begin{table}[H]
    \centering
    \caption{Comparison of first six frequencies for a cantilever gradient beam.}
      
 \begin{tabular}{c@{\hskip 0.01in}c@{\hskip 0.1in}c@{\hskip 0.1in}c@{\hskip 0.1in}c@{\hskip 0.1in}c@{\hskip 0.1in}c@{\hskip 0.1in}c@{\hskip 0.1in}}
     \\ \hline
 & N & $\bar{\omega}_{1}$    & $\bar{\omega}_{2}$  & $\bar{\omega}_{3}$& $\bar{\omega}_{4}$ &$\bar{\omega}_{5}$ &$\bar{\omega}_{6}$
\\ \hline 

   \multirow{3}{*}{} &  \multicolumn{1}{c}{}  & \multicolumn{1}{c}{} &  \multicolumn{1}{c}{$(\frac{g_{1}}{L}=0.1$,}& \multicolumn{1}{c}{$\frac{g_{2}}{L}=0.05)$}& \multicolumn{1}{c}{}& \multicolumn{1}{c}{} \\    
\cline{4-5}
  
&$5$& 4.3401   &  25.5081 &  65.3349  & --- & ---  & ---   \\

&$7$&  4.4726  &   27.9495  &  81.6317  &  272.0331  & 272.0331 &--- \\ 

&$9$&  4.3794   &   28.4598 &  89.6828  &  212.6240 &  212.6240 &  1286.9187 \\ 

 &$11$&4.4826   &   30.0222 &  91.9593  &  205.0360 &  559.2756 &  559.2756 \\

&$13$& 4.5560   &   30.4765 &  94.4327  &  214.8785 &  397.4738 &  627.8038 \\

&$15$  &4.5506   &   30.2841 &  94.1460  &  212.1549 &  408.0297 &  729.5127   \\ 
&$17$  	 &  4.5381   &   30.1294 &  93.7373  &  210.7456 &  403.3498 &  697.6757  \\ 
&$19$  	&4.5321   &   30.0660 &  93.5528  &  210.1586 &  402.3776 &  698.0781   \\
&$21$  	& 4.5301   &   30.0459 &  93.4918  &  209.9751 &  401.9982 &  696.9199   \\  
&Analyt. & 4.5320         &  30.0400 &  93.4723  &  209.9177 &  401.8808 &  696.6921	 \\ \hline 
   \multirow{3}{*}{} &  \multicolumn{1}{c}{}  & \multicolumn{1}{c}{} &  \multicolumn{1}{c}{$(\frac{g_{1}}{L}=0.15,$}& \multicolumn{1}{c}{$\frac{g_{2}}{L}=0.1)$}& \multicolumn{1}{c}{}& \multicolumn{1}{c}{} \\    
\cline{4-5}   

&$5$& 5.1344 &  48.7656   & 48.7656  & --- & ---  & ---   \\

&$7$&  5.1339 &  33.8646  &  91.9452  &  318.4791 & 602.5529 &--- \\ 

&$9$& 5.6158 &  41.1355 &  149.6005 & 325.1466 & 667.0974 & 864.7130 \\ 

& $11$& 5.4907 	 & 38.9018 &  132.2349  & 321.3672  &  855.4734 &  855.4734 \\
$0.1$&$13$  	& 5.4463 & 38.2281 &   130.4734 &   331.3741  & 693.3906 &  1115.4421 \\ 
&$15$  	 &  5.4394       & 38.1358 & 130.0263 & 326.6949 &  701.9729 & 1393.5781 \\ 
&$17$  	 &5.4391  &    38.1273 & 129.9964  &  326.8027 & 698.8957 &   1330.1310  \\
&$19$  	 &  5.4393   &   38.1268  & 129.9941 & 326.7796 & 699.1914 &  1341.5299  \\ 
&$21$  	 & 5.4323  &  38.1255  & 129.9935  & 326.7793 &  699.1701 & 1340.1697 \\ 

&Analyt. & 5.4324  &  38.0950  & 129.9835  & 326.7750 &  699.1683 & 1340.2756 \\ \hline 

        \end{tabular}
    \label{Freq_cantilever_beam}
\end{table}

\setlength{\extrarowheight}{.5em}
\begin{table}[H]
    \centering
    \caption{Comparison of first six elastic frequencies for a free-free gradient beam.}
      
 \begin{tabular}{c@{\hskip 0.01in}c@{\hskip 0.1in}c@{\hskip 0.1in}c@{\hskip 0.1in}c@{\hskip 0.1in}c@{\hskip 0.1in}c@{\hskip 0.1in}c@{\hskip 0.1in}}
     \\ \hline
& N & $\bar{\omega}_{1}$    & $\bar{\omega}_{2}$  & $\bar{\omega}_{3}$& $\bar{\omega}_{4}$ &$\bar{\omega}_{5}$ &$\bar{\omega}_{6}$
\\ \hline 
   \multirow{3}{*}{} &  \multicolumn{1}{c}{}  & \multicolumn{1}{c}{} &  \multicolumn{1}{c}{$(\frac{g_{1}}{L}=0.1$,}& \multicolumn{1}{c}{$\frac{g_{2}}{L}=0.05)$}& \multicolumn{1}{c}{}& \multicolumn{1}{c}{} \\    
\cline{4-5}
  
&$5$&   18.0353 &  --- &  --- & --- & ---  & ---   \\

&$7$ &22.5762 &    59.5169 &  104.9781  & ---  &--- &--- \\  

&$9$&   23.1959 &  71.3456 &   177.7259 &  240.5436 &  271.3764 & --- \\ 

 &$11$& 23.3713       &  71.5657 &   159.0272  &  342.6494 &  584.0780 & 633.2793   \\

&$13$&  23.4183 	 &  71.9297 &  161.3527  & 304.6324  &  507.0495 &  1243.2544 \\

&$15$  & 23.4335       &  71.9950 &  161.3431 &  309.6856 &  543.5879 &   835.1189   \\ 
&$17$  	 &   23.4382       &  72.0158 &  161.3771 &  309.2322  & 538.0707 &   885.8146\\ 
&$19$  	& 23.4393          &  72.0217 &  161.3850 &  309.2702 &  538.5042 &   875.6494   \\
&$21$  	&23.4398          &  72.0231 &  161.3869 &  309.2707 &  538.4716 &   876.7202 \\  
&Analyt. &23.4398          &  72.0235 &  161.3873 &  309.2713 &  538.4729 &   876.6273  \\ \hline 
   \multirow{3}{*}{} &  \multicolumn{1}{c}{}  & \multicolumn{1}{c}{} &  \multicolumn{1}{c}{$(\frac{g_{1}}{L}=0.15,$}& \multicolumn{1}{c}{$\frac{g_{2}}{L}=0.1)$}& \multicolumn{1}{c}{}& \multicolumn{1}{c}{} \\    
\cline{4-5}   

&$5$&   18.3813  &  ---   & ---  & --- & ---  & ---   \\

&$7$& 23.7026 &  64.6455  & 123.6199  &  ---  & --- &--- \\ 

&$9$&  24.2394 &  82.0112 &   233.2016 & 315.7059 & 384.1724 &--- \\ 

& $11$&  24.3203 	 &  81.4475 &  201.4857  & 497.4937  &   1035.1812 &  --- \\
&$13$  	&24.3256 &  81.6035 &   203.8481 &  433.2935  &  810.0301 &   2392.5067\\ 
&$15$  	 & 24.3258     & 81.5935 & 203.6459 &  440.1209 &  866.9113 &  1482.3796 \\ 
&$17$  	 & 24.3259     & 81.5927 & 203.6495 &  439.3965 &  858.9223 &  1570.1713 \\
&$19$  	 &  24.3256   &  81.5927  & 203.6493 & 439.4347 & 859.5209 &   1554.4556  \\ 
&$21$  	 &  24.3239  &  81.5928  & 203.6493 & 439.4332 & 859.4868 &   1556.0376  \\ 

&Analyt. & 24.3230   &  81.5930  & 203.6489 & 439.4334 & 859.4885 &   1555.9236	 \\ \hline

        \end{tabular}
    \label{Freq_Free_Free_beam}
\end{table}

Excellent fit with the analytical solutions is noticed in the fundamental frequencies obtained using the proposed element with fewer number of grid points. This consistency is maintained for all the boundary conditions and length scale parameters. Hence, based on the above findings it can be stated that the present element can be efficiently applied for free vibration analysis of second strain gradient Euler-Bernoulli beam for any choice of boundary conditions and $g/L$ values.
 
\subsection{Stability analysis of gradient elastic beams}

In the earlier sections the efficiency of the  proposed beam element was verified for static and free vibration analysis of gradient elastic beams. Here, we validate the applicability of the element for stability analysis of second strain gradient Euler-Bernoulli beam under different support conditions. The DQ results are compared with the analytical values obtained in Appendix-I for different support conditions. The convergence of critical buckling load for a simply supported beam obtained for $g_{1}/L=0.1$, $g_{2}/L=0.05$ and $g_{1}/L=0.15$, $g_{2}/L=0.1$  are shown in Table \ref{Pcr_ss}. It can be noticed that the convergence of buckling load is rapid and approaches to analytical values with 15 grid points for all the $g/L$ values. Similar convergence behaviour is noticed in Table \ref{Pcr_All_Bcs}, for a clamped, cantilever and propped cantilever beam. Hence, these observations validate the effectiveness of the proposed beam element for buckling analysis of second strain gradient elastic prismatic beams. 

\setlength{\extrarowheight}{.5em}
\begin{table}
    \centering
    \caption{Comparison of normalized buckling load for a simply supported gradient beam.}

\begin{tabular}{|c@{\hskip 0.1in}|c@{\hskip 0.1in}|c@{\hskip 0.1in}|} 
\hline
$ \text{N}$ & $\frac{g_{1}}{L}=0.1,\frac{g_{2}}{L}=0.05$	& $\frac{g_{1}}{L}=0.15,\frac{g_{2}}{L}=0.1$   \\
 \hline 
 $5$	 & 10.6471       &   11.7216    \\  
 $7$ &    11.0387&  13.2749   \\ 
$9$	&   11.0386 &   13.4545  \\
 $11$	 &  11.2035       &  13.0837   \\  
 $13$	 &  11.1359       &  13.0163   \\ 
 $15$	 &  11.0302       &  13.0091   \\ 
 $17$	 &  10.9867       &  13.0085   \\ 
 $19$	 &  10.9737       &  13.0085   \\ 
 $21$	 &  10.9705       &  13.0085   \\ 
Analytical  &  10.9704       &  13.0084 	 \\  \hline

\end{tabular}
    \label{Pcr_ss}
\end{table}

\begin{table}[H]
    \centering
    \caption{Comparison of normalized buckling load for a clamped, cantilever and propped-cantilever gradient beams.}

\begin{tabular}{|c@{\hskip 0.15in}|c@{\hskip 0.15in}c@{\hskip 0.1in}|c@{\hskip 0.1in}c@{\hskip 0.1in}|c@{\hskip 0.15in}c@{\hskip 0.15in}|}
\hline
\multirow{3}{*}{N} & \multicolumn{2}{c|}{Clamped} & %
 \multicolumn{2}{c|}{Cantilever} & %
 \multicolumn{2}{c|}{Propped cantilever}\\
\cline{2-7}
 & $\frac{g_{1}}{L}=0.1$ & $\frac{g_{1}}{L}=0.15$ & $\frac{g_{1}}{L}=0.1$ & $\frac{g_{1}}{L}=0.15$ & $\frac{g_{1}}{L}=0.1$ & $\frac{g_{1}}{L}=0.15$\\
  & $\frac{g_{2}}{L}=0.05$ & $\frac{g_{2}}{L}=0.1$ & $\frac{g_{2}}{L}=0.05$ & $\frac{g_{2}}{L}=0.1$ & $\frac{g_{2}}{L}=0.05$ & $\frac{g_{2}}{L}=0.1$\\

\cline{1-7}
 $5$	    &   93.8695  & 157.5518   &    3.0490  & 3.8694  &   38.1265  & 84.9298 \\
 $7$ & 78.4093 &  172.4970 &  3.3739  &   3.7910  &  29.0109  &   66.7547 \\ 
$9$	&   87.8053 &   221.4184 &   3.1775 &  4.0903  &   31.3016 &    63.4955 \\
 $11$  &  94.4424& 240.1186    & 3.1755 &   4.0818  & 33.1909 &56.1652    \\  
  $13$	 & 98.8326  & 230.7690       &  3.2643 & 4.0608  &  33.3466 &  54.2149  \\
 $15$	 &  98.4481 & 230.0158    &   3.2759 & 4.0571 &   32.7474 & 53.9577 \\
$17$	    &  97.6336 & 229.4688      & 3.2707 &  4.0567 & 32.4233 &  53.9347\\
$19$	    & 97.2638  & 229.9457     &  3.2674  & 4.0571 &  32.3046  &  53.9334 \\ 
$21$	    & 97.1446  & 229.9458     &  3.2663  & 4.0566 &  32.2688  &  53.9334 \\ 
Analytical  & 97.1445  & 229.9456     &  3.2661  & 4.0565 &  32.2686  &  53.9331   \\
 \hline

\end{tabular}

    \label{Pcr_All_Bcs}
\end{table}

\section{Conclusion}

Variational formulation for a second strain gradient Euler-Bernoulli beam theory was presented for the first time, and the governing equation and associated classical and non-classical boundary conditions were obtained. A novel differential quadrature beam element was proposed to solve a eight order partial differential equation which governs the second strain gradient Euler-Bernoulli beam theory. The element was formulated using the strong form of the governing equation in conjunction with the Lagrange interpolation functions. A new way to account for the non-classical boundary conditions associated with the gradient elastic beam was introduced. The efficiency and accuracy of the proposed element was established through application to static, free vibration and stability  analysis of gradient elastic beams for different support conditions and length scale parameters.

\section*{APPENDIX}

\section*{Analytical solutions for second strain gradient Euler-Bernoulli beam}

In this section we obtain the analytical solutions for bending, free vibration and stability analysis of second strain gradient Euler-Bernoulli beam for different support conditions and length scale parameters.

\subsection*{Bending analysis}
Let us consider a beam of length \textit{L} subjected to a uniformly distributed load \textit{q}. To obtain the static deflections of the second gradient elastic Euler-Bernoulli beam which is governed by Equation (\ref{EOM_beam}), we assume a solution of the form

 \begin{align*}  \label{exact_disp}       
w(x)= c_{1}+c_{2}x+c_{3}x^{2}+c_{4}x^{3}+c_{5}e^{n_{1}x}+
c_{6}e^{n_{2}x}+c_{7}e^{m_{1}x}+c_{8}e^{m_{2}x}-\frac{qx^{4}}{24EI} \tag{a1}
\end{align*} 

where 
 \begin{align*}         
n_{1}=\sqrt{\frac{g_{1}^{2}+\sqrt{g_{1}^{4}-4g_{2}^{4}}}{2g_{2}^{4}}}, \quad
n_{2}=-\sqrt{\frac{g_{1}^{2}+\sqrt{g_{1}^{4}-4g_{2}^{4}}}{2g_{2}^{4}}}, \quad \\ m_{1}=\sqrt{\frac{g_{1}^{2}-\sqrt{g_{1}^{4}-4g_{2}^{4}}}{2g_{2}^{4}}}, \quad
m_{2}=-\sqrt{\frac{g_{1}^{2}-\sqrt{g_{1}^{4}-4g_{2}^{4}}}{2g_{2}^{4}}}, \quad
\end{align*}

\noindent The constants $c_{1}-c_{8}$ are determined with the aid of boundary conditions listed in Equation (\ref{eq:BC_Cl_Beam}) and (\ref{eq:BC_NCl_Beam}). After applying the boundary conditions the system of equations are expressed as:

\begin{align*}        
[K]\{\delta\}=\{f\} \tag{a2}
\end{align*}

Here $K$ is the coefficient matrix, $f$ is the vector corresponding to the force and $\{\delta\}=\{c_{1},c_{2},c_{3},c_{4},c_{5},c_{6},c_{7},c_{8}\}$ is the unknown constant vector to be determined. Once the unknown constants are determined then the displacement solution is obtained from the Equation (\ref{exact_disp}). The slope, curvature and triple derivative of displacement at any point along the length of the beam can be obtained by performing the first, second and third derivatives of the deflection respectively. The shear force, bending moment, double moment and triple moment are obtained by substituting the Equation (\ref{exact_disp}) in Equations(\ref{eq:BC_Cl_Beam}) and (\ref{eq:BC_NCl_Beam}). To have real and positive roots $g_{1}/g_{2} > \sqrt{2}$ is assumed in the present analysis.  The following are the list of simultaneous equations to determine the unknown coefficients for different boundary conditions: \\

\noindent (a) Simply supported beam :
\begin{align*}
[K]=
\begin{bmatrix}
1 & 0 & 0 & 0 & 1 & 1 & 1 & 1\\
1 & L & L^2 & L^3 & e^{m_{1}L} & e^{m_{2}L} & e{n_{1}L} & e^{n_{2}L}\\
0 & 0 & 2 & 0 & a_{11} & a_{12}  &
a_{13} & a_{14}\\
    0 & 0 & 2 & 6L & b_{11} & b_{12} &
b_{13} & b_{14}\\
     0 & 0 & 2 & 0 & m_{1}^{2} & m_{2}^{2} & n_{1}^{2}  & n_{2}^{2} \\
     0 & 0 & 2 & 6L & m_{1}^{2}e^{m_{1}L} & m_{2}^{2}e^{m_{2}L} & n_{1}^{2}e^{n_{1}L} & n_{2}^{2}e^{n_{2}L}\\
     0 & 0 & 0 & 6 & m_{1}^{3} & m_{2}^{3} & n_{1}^{3} & n_{2}^{3}\\
     0 & 0 & 0 & 6 & m_{1}^{3}e^{m_{1}L} & m_{2}^{3}e^{m_{2}L} & n_{1}^{3}e^{n_{1}L} & n_{2}^{3}e^{n_{2}L}\\
     \end{bmatrix}, \,\,\, \{f\}=\begin{Bmatrix}
 0 \\ 
-qL^{4}/24EI \\ 
g_{1}^{2}q/EI\\ 
g_{1}^{2}q/EI-qL^{2}/2EI\\ 
 0 \\ 
-qL^{2}/2EI\\ 
0\\
-qL/EI\\
\end{Bmatrix} 
\end{align*}
\noindent where, \\
\noindent $ a_{11}= m_{1}^{2}-g_{1}^{2}m_{1}^{4}+g_{2}^{4}m_{1}^{6},\,\,\, \quad a_{12}= m_{2}^{2}-g_{1}^{2}m_{2}^{4}+g_{2}^{4}m_{2}^{6},\\ a_{13}=n_{1}^{2}-g_{1}^{2}n_{1}^{4}+g_{2}^{4}n_{1}^{6}\,\,, \,\,\,\,\,a_{14}= n_{2}^{2}-g_{1}^{2}n_{2}^{4}+g_{2}^{4}n_{2}^{6}\\ \\
b_{11}= (m_{1}^{2}-g_{1}^{2}m_{1}^{4}+g_{2}^{4}m_{1}^{6})e^{m_{1}L}, \,\,\,b_{12}= (m_{2}^{2}-g_{1}^{2}m_{2}^{4}+g_{2}^{4}m_{2}^{6})e^{m_{2}L},
\\ b_{13}= (n_{1}^{2}-g_{1}^{2}n_{1}^{4}+g_{2}^{4}n_{1}^{6})e^{n_{1}L},\,\,\,\,\,\,\,\,b_{14}= (n_{2}^{2}-g_{1}^{2}n_{2}^{4}+g_{2}^{4}n_{2}^{6})e^{n_{2}L}
$ \\

\noindent (b) Cantilever beam :

\begin{align*}
[K]=
\begin{bmatrix}
1 & 0 & 0 & 0 & 1 & 1 & 1 & 1\\
0 & 0 & 0 & 6 & a_{21} &a_{22} & a_{23} &a_{24}\\
0 & 1 & 0 & 0 & m_{1} & m_{2}  &
n_{1} & n_{2}\\
0 & 0 & 2 & 6L & b_{11} & b_{12} &
b_{13} & b_{14}\\
0 & 0 & 2 & 0 & m_{1}^{2} & m_{2}^{2} & n_{1}^{2}  & n_{2}^{2} \\
0 & 0 & 0 & 6g_{1}^{2} & b_{21} & b_{22} &
b_{23} & b_{24}\\
0 & 0 & 0 & 6 & m_{1}^{3} & m_{2}^{3} & n_{1}^{3} & n_{2}^{3}\\
0 & 0 & 0 & 0 & c_{11} & c_{12} & c_{13} & c_{14}\\
     \end{bmatrix}, \,\,\, \{f\}=\begin{Bmatrix}
 0 \\ 
-qL/EI \\ 
0\\
g_{1}^{2}q/EI-qL^{2}/2EI\\ 
0\\
-g_{1}^{2}Lq/EI\\ 
 0\\
0\\
\end{Bmatrix} 
\end{align*}

\noindent where, \\
\noindent $a_{21}=(m_{1}^{3}-g_{1}^{2}m_{1}^{5}+g_{2}^{4}m_{1}^{7})e^{m_{1}L},\,\,\,
a_{22}=(m_{2}^{3}-g_{1}^{2}m_{2}^{5}+g_{2}^{4}m_{2}^{7})e^{m_{2}L},\,\,\,
\\a_{23}=(n_{1}^{3}-g_{1}^{2}n_{1}^{5}+g_{2}^{4}n_{1}^{7})e^{n_{1}L},\,\,\,
a_{24}=(n_{2}^{3}-g_{1}^{2}n_{2}^{5}+g_{2}^{4}n_{2}^{7})e^{n_{2}L},\,\,\,\\
\\b_{21}=(g_{1}^{2}m_{1}^{3}-g_{2}^{4}m_{1}^{5})e^{m_{1}L},\,\,\,
b_{22}=(g_{1}^{2}m_{2}^{3}-g_{2}^{4}m_{2}^{5})e^{m_{2}L},\,\,\,\\
b_{23}=(g_{1}^{2}n_{1}^{3}-g_{2}^{4}n_{1}^{5})e^{n_{1}L},\,\,\,
b_{24}=(g_{1}^{2}n_{2}^{3}-g_{2}^{4}n_{2}^{5})e^{n_{2}L}\\ \\
c_{11}=g_{2}^{4}m_{1}^{4}e^{m_{1}L},\,\,\,c_{12}=g_{2}^{4}m_{2}^{4}e^{m_{2}L},\,\,\,c_{13}=g_{2}^{4}n_{1}^{4}e^{n_{1}L},\,\,\,c_{14}=g_{2}^{4}n_{2}^{4}e^{n_{2}L},\,\,\,
$ \\

\noindent (c) Clamped beam :

\begin{align*}
[K]=
\begin{bmatrix}
1 & 0 & 0 & 0 & 1 & 1 & 1 & 1\\
1 & L & L^{2} & L^{3} & e^{m_{1}L} & e^{m_{2}L} & e^{n_{1}L} &e^{n_{2}L}\\
0 & 1 & 0 & 0 & m_{1} & m_{2}  & n_{1} & n_{2} \\
0 & 1 & 2L & 3L^{2} & m_{1}e^{m_{1}L} & m_{2}e^{m_{2}L} & n_{1}e^{n_{1}L} & n_{2}e^{n_{2}L}\\ 
0 & 0 & 2 & 0 & m_{1}^{2} & m_{2}^{2} & n_{1}^{2}  & n_{2}^{2} \\
0 & 0 & 2 & 6L & m_{1}^{2}e^{m_{1}L} & m_{2}^{2}e^{m_{2}L} & n_{1}^{2}e^{n_{1}L} & n_{2}^{2}e^{n_{2}L}\\
0 & 0 & 0 & 6 & m_{1}^{3} & m_{2}^{3} & n_{1}^{3}  & n_{2}^{3} \\
0 & 0 & 0 & 6 &  m_{1}^{3}e^{m_{1}L} & m_{3}^{2}e^{m_{2}L} & n_{1}^{3}e^{n_{1}L} & n_{2}^{3}e^{n_{2}L}\\
  \end{bmatrix}, \,\,\, \{f\}=\begin{Bmatrix}
 0 \\ 
-qL^{4}/24EI \\ 
0\\
-qL^{3}/6EI \\ 
0\\
-qL^{2}/2EI\\ 
 0\\
-qL/EI\\ 
\end{Bmatrix}
\end{align*}

\subsection*{Free vibration analysis}

To obtain the natural frequencies of the second gradient elastic Euler-Bernoulli beam which is governed by Equation (\ref{EOM_beam}), we assume a solution of the form

\begin{align*}         
w(x,t)=\bar{w}(x){e}^{i\omega{t}} \tag{b1}
\end{align*}

\noindent substituting the above solution in the governing equation (\ref{EOM_beam}), we get

\begin{align*}         
\bar{w}^{iv}-g_{1}^{2}\bar{w}^{vi}+g_{2}^{4}\bar{w}^{viii}-\frac{\omega^{2}}{\beta^{2}}\bar{w}=0 \tag{b2}
\end{align*}

\noindent here, $\beta^{2}=EI/m$, and the above equation has the solution of type

\begin{align*}         
\bar{w}(x)=\sum_{j=1}^{8}c_{i}{e}^{k_{i}x} \tag{b3}
\end{align*}

\noindent where, $c_{i}$ are the constants of integration which are determined through boundary conditions and the $k_{i}$ are the roots of the characteristic equation 

\begin{align*}         
{k}^{iv}-g_{1}^{2}{k}^{vi}+g_{2}^{4}{k}^{viii}-\frac{\omega^{2}}{\beta^{2}}=0 \tag{b4}
\end{align*}

After applying the boundary conditions listed in Section \ref{Sg_formulation}, we get,

\begin{align*}         
[F(\omega)]\{C\}=\{0\} \tag{b5}
\end{align*}

For non-trivial solution, following condition should be satisfied

\begin{align*}         
det[F(\omega)]=0 \tag{b6}
\end{align*}

The above frequency equation renders all the natural frequencies for a second strain gradient Euler-Bernoulli beam. The following are the frequency equations for different boundary conditions:\\

\noindent (a) Simply supported beam :
\begin{align*}\label{exact_freq_ss_EQ}
[F(\omega)]=\begin{bmatrix}
1 & 1 & 1 & 1 & 1 & 1& 1 & 1\\
{e}^{(k_{1}L)} & {e}^{(k_{2}L)} & {e}^{(k_{3}L)} & {e}^{(k_{4}L)} & {e}^{(k_{5}L)} & {e}^{(k_{6}L)}& {e}^{(k_{7}L)} & {e}^{(k_{8}L)} \\
{k_{1}}^2 & {k_{2}}^2 & {k_{3}}^2 & {k_{4}}^2 & {k_{5}}^2 & {k_{6}}^2& {k_{7}}^2 & {k_{8}}^2\\ 
t_{1} &t_{2}&t_{3}&t_{4}&t_{5}&t_{6}&t_{7}&t_{8}\\ 
t_{1}{e}^{(k_{1}L)} &t_{2}{e}^{(k_{2}L)}&t_{3}{e}^{(k_{3}L)}&t_{4}{e}^{(k_{4}L)}&t_{5}{e}^{(k_{5}L)}&t_{6}{e}^{(k_{6}L)}&t_{7}{e}^{(k_{7}L)}&t_{8}{e}^{(k_{8}L)}\\
k_{1}^{3} & k_{2}^{3} & k_{3}^{3} & k_{4}^{3} & k_{5}^{3} & k_{6}^{3}& k_{7}^{3} & k_{8}^{3}  \\
k_{1}^{3}{e}^{(k_{1}L)} & k_{2}^{3}{e}^{(k_{2}L)} & k_{3}^{3}{e}^{(k_{3}L)} & k_{4}^{3}{e}^{(k_{4}L)} & k_{5}^{3}{e}^{(k_{5}L)} & k_{6}^{3}{e}^{(k_{6}L)} &  k_{7}^{3}{e}^{(k_{7}L)} & k_{8}^{3}{e}^{(k_{8}L)}\\
\end{bmatrix}
\end{align*}

\noindent (b) Cantilever beam :
\begin{align*}
[F(\omega)]=\begin{bmatrix}
1 & 1 & 1 & 1 & 1 & 1& 1 & 1\\
{k_{1}} & {k_{2}} & {k_{3}} & {k_{4}} & {k_{5}} & {k_{6}}& {k_{7}} & {k_{8}}\\ 
{k_{1}}^2 & {k_{2}}^2 & {k_{3}}^2 & {k_{4}}^2 & {k_{5}}^2 & {k_{6}}^2& {k_{7}}^2 & {k_{8}}^2\\ 
{k_{1}}^3 & {k_{2}}^3 & {k_{3}}^3 & {k_{4}}^3 & {k_{5}}^3 & {k_{6}}^3& {k_{7}}^3 & {k_{8}}^3\\ 
p_{1}{e}^{(k_{1}L)} &p_{2}{e}^{(k_{2}L)}&p_{3}{e}^{(k_{3}L)}&p_{4}{e}^{(k_{4}L)}&p_{5}{e}^{(k_{5}L)}&p_{6}{e}^{(k_{6}L)}&p_{7}{e}^{(k_{7}L)}&p_{8}{e}^{(k_{8}L)}\\ 
r_{1}{e}^{(k_{1}L)} &r_{2}{e}^{(k_{2}L)}&r_{3}{e}^{(k_{3}L)}&r_{4}{e}^{(k_{4}L)}&r_{5}{e}^{(k_{5}L)}&r_{6}{e}^{(k_{6}L)}&r_{7}{e}^{(k_{7}L)}&r_{8}{e}^{(k_{8}L)}\\ 
q_{1}{e}^{(k_{1}L)} &q_{2}{e}^{(k_{2}L)}&q_{3}{e}^{(k_{3}L)}&q_{4}{e}^{(k_{4}L)}&q_{5}{e}^{(k_{5}L)}&q_{6}{e}^{(k_{6}L)}&q_{7}{e}^{(k_{7}L)}&q_{8}{e}^{(k_{8}L)}\\ 
k_{1}^{4}{e}^{(k_{1}L)} & k_{2}^{4}{e}^{(k_{2}L)} & k_{3}^{4}{e}^{(k_{3}L)} & k_{4}^{4}{e}^{(k_{4}L)} & k_{5}^{4}{e}^{(k_{5}L)} & k_{6}^{4}{e}^{(k_{6}L)} &  k_{7}^{4}{e}^{(k_{7}L)} & k_{8}^{4}{e}^{(k_{8}L)}\\
\end{bmatrix} 
\end{align*}

\noindent (c) clamped beam :
\begin{align*}
[F(\omega)]=\begin{bmatrix}
1 & 1 & 1 & 1 & 1 & 1& 1 & 1\\
{e}^{(k_{1}L)} & {e}^{(k_{2}L)} & {e}^{(k_{3}L)} & {e}^{(k_{4}L)} & {e}^{(k_{5}L)} & {e}^{(k_{6}L)}& {e}^{(k_{7}L)} & {e}^{(k_{8}L)} \\
{k_{1}} & {k_{2}} & {k_{3}} & {k_{4}} & {k_{5}} & {k_{6}}& {k_{7}} & {k_{8}}\\ 
{k_{1}}{e}^{(k_{1}L)} & {k_{2}}{e}^{(k_{2}L)} & {k_{3}}{e}^{(k_{3}L)} & {k_{4}}{e}^{(k_{4}L)} & {k_{5}}{e}^{(k_{5}L)} & {k_{6}}{e}^{(k_{6}L)}& {k_{7}}{e}^{(k_{7}L)} & {k_{8}}{e}^{(k_{8}L)} \\
{k_{1}}^2 & {k_{2}}^2 & {k_{3}}^2 & {k_{4}}^2 & {k_{5}}^2 & {k_{6}}^2& {k_{7}}^2 & {k_{8}}^2\\ 
k_{1}^{2}{e}^{(k_{1}L)} & k_{2}^{2}{e}^{(k_{2}L)} & k_{3}^{2}{e}^{(k_{3}L)} & k_{4}^{2}{e}^{(k_{4}L)} & k_{5}^{2}{e}^{(k_{5}L)} & k_{6}^{2}{e}^{(k_{6}L)} &  k_{7}^{2}{e}^{(k_{7}L)} & k_{8}^{2}{e}^{(k_{8}L)}\\
{k_{1}}^3 & {k_{2}}^3 & {k_{3}}^3 & {k_{4}}^3 & {k_{5}}^3 & {k_{6}}^3& {k_{7}}^3 & {k_{8}}^3\\ 
k_{1}^{3}{e}^{(k_{1}L)} & k_{2}^{3}{e}^{(k_{2}L)} & k_{3}^{3}{e}^{(k_{3}L)} & k_{4}^{3}{e}^{(k_{4}L)} & k_{5}^{3}{e}^{(k_{5}L)} & k_{6}^{3}{e}^{(k_{6}L)} &  k_{7}^{3}{e}^{(k_{7}L)} & k_{8}^{3}{e}^{(k_{8}L)}\\
\end{bmatrix} 
\end{align*}

\noindent (d) Free-free beam :
\begin{align*}
[F(\omega)]=
\begin{bmatrix}
p_{1} &p_{2}&p_{3}&p_{4}&p_{5}&p_{6}&p_{7}&p_{8}\\ 
r_{1} &r_{2}&r_{3}&r_{4}&r_{5}&r_{6}&r_{7}&r_{8}\\ 
q_{1} &q_{2}&q_{3}&p_{4}&q_{5}&q_{6}&q_{7}&q_{8}\\ 
k_{1}^{4} & k_{2}^{4} & k_{3}^{4} & k_{4}^{4} & k_{5}^{4} & k_{6}^{4}&  k_{7}^{4} & k_{8}^{4}\\
p_{1}{e}^{(k_{1}L)} &p_{2}{e}^{(k_{2}L)}&p_{3}{e}^{(k_{3}L)}&p_{4}{e}^{(k_{4}L)}&p_{5}{e}^{(k_{5}L)}&p_{6}{e}^{(k_{6}L)}&p_{7}{e}^{(k_{7}L)}&p_{8}{e}^{(k_{8}L)}\\ 
r_{1}{e}^{(k_{1}L)} &r_{2}{e}^{(k_{2}L)}&r_{3}{e}^{(k_{3}L)}&r_{4}{e}^{(k_{4}L)}&r_{5}{e}^{(k_{5}L)}&r_{6}{e}^{(k_{6}L)}&r_{7}{e}^{(k_{7}L)}&r_{8}{e}^{(k_{8}L)}\\ 
q_{1}{e}^{(k_{1}L)} &q_{2}{e}^{(k_{2}L)}&q_{3}{e}^{(k_{3}L)}&p_{4}{e}^{(k_{4}L)}&q_{5}{e}^{(k_{5}L)}&q_{6}{e}^{(k_{6}L)}&s_{7}{e}^{(k_{7}L)}&q_{8}{e}^{(k_{8}L)}\\ 
k_{1}^{4}{e}^{(k_{1}L)} & k_{2}^{4}{e}^{(k_{2}L)} & k_{3}^{4}{e}^{(k_{3}L)} & k_{4}^{4}{e}^{(k_{4}L)} & k_{5}^{4}{e}^{(k_{5}L)} & k_{6}^{4}{e}^{(k_{6}L)} &  k_{7}^{4}{e}^{(k_{7}L)} & k_{8}^{4}{e}^{(k_{8}L)}\\
\end{bmatrix} 
\end{align*}

\noindent Where,  \\

\noindent $t_{1}=(-g_{1}^{2}k_{1}^{4}+g_{2}^{4}k_{1}^{6}),\quad  t_{2}=(-g_{1}^{2}k_{2}^{4}+g_{2}^{4}k_{2}^{6}) ,\quad t_{3}=(-g_{1}^{2}k_{3}^{4}+g_{2}^{4}k_{3}^{6})$ \\ 
$t_{4}=(-g_{1}^{2}k_{4}^{4}+g_{2}^{4}k_{4}^{6}) ,\quad t_{5}=(-g_{1}^{2}k_{5}^{4}+g_{2}^{4}k_{5}^{6}) \quad t_{6}=(-g_{1}^{2}k_{6}^{4}+g_{2}^{4}k_{6}^{6})$ \\
$t_{7}=(-g_{1}^{2}k_{7}^{4}+g_{2}^{4}k_{7}^{6}) ,\quad t_{8}=(-g_{1}^{2}k_{8}^{4}+g_{2}^{4}k_{8}^{6})$ \\

\noindent $p_{1}=(k_{1}^{3}-g_{1}^{2}{k_{1}}^{5}+g_{2}^{4}{k_{1}}^{7}) ,\quad  p_{2}=(k_{2}^{3}-g_{1}^{2}{k_{2}}^{5}+g_{2}^{4}{k_{2}}^{7})\quad \\ p_{3}=(k_{3}^{3}-g_{1}^{2}{k_{3}}^{5}+g_{2}^{4}{k_{3}}^{7})$ , \,\, 
$p_{4}=(k_{4}^{3}-g_{1}^{2}{k_{4}}^{5}+g_{2}^{4}{k_{4}}^{7}) \\ \quad p_{5}=(k_{5}^{3}-g_{1}^{2}{k_{5}}^{5}+g_{2}^{4}{k_{5}}^{7}), \quad p_{6}=(k_{6}^{3}-g_{1}^{2}{k_{6}}^{5}+g_{2}^{4}{k_{6}}^{7}) \\ p_{7}=(k_{7}^{3}-g_{1}^{2}{k_{7}}^{5}+g_{2}^{4}{k_{7}}^{7}), \quad p_{8}=(k_{8}^{3}-g_{1}^{2}{k_{8}}^{5}+g_{2}^{4}{k_{8}}^{7})$ \\

\noindent $r_{1}=(k_{1}^{2}-g_{1}^{2}{k_{1}}^{4}+g_{2}^{4}{k_{1}}^{6}) ,\quad  r_{2}=(k_{2}^{2}-g_{1}^{2}{k_{2}}^{4}+g_{2}^{4}{k_{2}}^{6}) \quad \\ r_{3}=(k_{3}^{2}-g_{1}^{2}{k_{3}}^{4}+g_{2}^{4}{k_{3}}^{6})$ , \,\, 
$r_{4}=(k_{4}^{2}-g_{1}^{2}{k_{4}}^{4}+g_{2}^{4}{k_{4}}^{6})  \\ \quad r_{5}=(k_{5}^{2}-g_{1}^{2}{k_{5}}^{4}+g_{2}^{4}{k_{5}}^{6}), \quad r_{6}=(k_{6}^{2}-g_{1}^{2}{k_{6}}^{4}+g_{2}^{4}{k_{6}}^{6}) \\ r_{7}=(k_{7}^{2}-g_{1}^{2}{k_{7}}^{4}+g_{2}^{4}{k_{7}}^{6}), \quad r_{8}=(k_{8}^{2}-g_{1}^{2}{k_{8}}^{4}+g_{2}^{4}{k_{8}}^{6})$ \\

\noindent $q_{1}=(g_{1}^{2}{k_{1}}^{3}-g_{2}^{4}{k_{1}}^{5}) ,\quad  q_{2}=(g_{1}^{2}{k_{2}}^{3}-g_{2}^{4}{k_{2}}^{5}) \quad \\ q_{3}=(g_{1}^{2}{k_{3}}^{3}-g_{2}^{4}{k_{3}}^{5})$ , \,\, 
$q_{4}=(g_{1}^{2}{k_{4}}^{3}-g_{2}^{4}{k_{4}}^{5})  \\ \quad q_{5}=(g_{1}^{2}{k_{5}}^{3}-g_{2}^{4}{k_{5}}^{5}), \quad q_{6}=(g_{1}^{2}{k_{6}}^{3}-g_{2}^{4}{k_{6}}^{5}) \\ q_{7}=(g_{1}^{2}{k_{7}}^{3}-g_{2}^{4}{k_{7}}^{5}), \quad q_{8}=(g_{1}^{2}{k_{8}}^{3}-g_{2}^{4}{k_{8}}^{5})$ \\

\subsection*{Stability analysis}

To obtain the buckling load for a second strain gradient Euler-Bernoulli beam which is governed by Equation(\ref{EOM_beam}), we assume a solution of the form

\begin{align*}         
w(x)=c_{1}+c_{2}x+c_{3}e^{m_{1}x}+c_{4}e^{m_{2}x}+c_{5}e^{m_{3}x}+
c_{6}e^{n_{1}x}+c_{7}e^{n_{2}x}+c_{8}e^{n_{3}x} \\ \tag{c1}
\end{align*}

\noindent where, $c_{i}$ are the constants of integration which are determined through boundary conditions and the $m_{1,2,3}$ and $n_{1,2,3}$ are the roots of the following characteristic equation:
\begin{align*}         
g_{2}^{4}s^{6}-g_{1}^{2}s^{4}+s^{2}+\frac{P}{EI}=0 \tag{c2}
\end{align*}

After applying the boundary conditions listed in section \ref{Sg_formulation}, we obtain,

\begin{align*}         
[\bar{G}(P)]\{C\}=\{0\}
\end{align*}

For non-trivial solution, following condition should be satisfied

\begin{align*}         
det[\bar{G}(P)]=0 \tag{c3}
\end{align*}

The above Eigenvalue problem renders the buckling load for a second strain gradient elastic Euler-Bernoulli beam. The following are the system equations for different boundary conditions.\\
 
\noindent (a) Simply supported beam :
\begin{align*}
[F(\omega)]=
\begin{bmatrix}
1 & 0 & 1 & 1 & 1 & 1& 1 & 1\\
1 & L & {e}^{(m_{1}L)} & {e}^{(m_{2}L)} & {e}^{(m_{3}L)} & {e}^{(n_{1}L)}& {e}^{(n_{2}L)} & {e}^{(n_{3}L)} \\
0 & 0 & {m_{1}}^2 & {m_{2}}^2 & {m_{3}}^2 & {n_{1}}^2& {n_{2}}^2 & {n_{3}}^2\\ 
0 & 0 & t_{3}&t_{4}&t_{5}&t_{6}&t_{7}&t_{8}\\ 
0 & 0 & t_{3}{e}^{(m_{1}L)}&t_{4}{e}^{(m_{2}L)}&t_{5}{e}^{(m_{3}L)}&t_{6}{e}^{(n_{1}L)}&t_{7}{e}^{(n_{2}L)}&t_{8}{e}^{(n_{3}L)}\\
0 & 0 & m_{1}^{3} & m_{2}^{3} & m_{3}^{3} & n_{1}^{3}& n_{2}^{3} & n_{3}^{3}  \\
0 & 0 & m_{1}^{3}{e}^{(m_{1}L)} & m_{2}^{3}{e}^{(m_{2}L)} & n_{1}^{3}{e}^{(n_{1}L)} & n_{2}^{3}{e}^{(n_{2}L)} &  n_{3}^{3}{e}^{(n_{3}L)} & n_{4}^{3}{e}^{(n_{4}L)}\\
\end{bmatrix} 
\end{align*}

\noindent (b) Cantilever beam :
\begin{align*}
[F(\omega)]=
\begin{bmatrix}
1 & 0 & 1 & 1 & 1 & 1& 1 & 1\\
0 & 1 & {m_{1}} & {m_{2}} & {m_{3}} & {n_{1}}& {n_{2}} & {n_{3}}\\ 
0 & 0 & {m_{1}}^2 & {m_{2}}^2 & {m_{3}}^2 & {n_{1}}^2& {n_{2}}^2 & {n_{3}}^2\\ 
0 & 0 & {m_{1}}^3 & {m_{2}}^3 & {m_{3}}^3 & {n_{1}}^3& {n_{2}}^3 & {n_{3}}^3\\ 
0 & 0 & p_{3}{e}^{(m_{1}L)}&p_{4}{e}^{(m_{2}L)}&p_{5}{e}^{(m_{3}L)}&p_{6}{e}^{(n_{1}L)}&p_{7}{e}^{(n_{2}L)}&p_{8}{e}^{(n_{3}L)}\\ 
0 & 0 & r_{3}{e}^{(m_{1}L)}&r_{4}{e}^{(m_{2}L)}&r_{5}{e}^{(m_{3}L)}&r_{6}{e}^{(n_{1}L)}&r_{7}{e}^{(n_{2}L)}&r_{8}{e}^{(n_{3}L)}\\ 
0 & 0 & s_{3}{e}^{(m_{1}L)}&p_{4}{e}^{(m_{2}L)}&s_{5}{e}^{(m_{3}L)}&s_{6}{e}^{(n_{1}L)}&s_{7}{e}^{(n_{2}L)}&s_{8}{e}^{(n_{3}L)}\\ 
0 & 0 & m_{1}^{4}{e}^{(m_{1}L)} & m_{2}^{4}{e}^{(_{2}L)} & m_{3}^{4}{e}^{(m_{3}L)} & n_{1}^{4}{e}^{(n_{1}L)} &  n_{2}^{4}{e}^{(n_{2}L)} & n_{3}^{4}{e}^{(n_{3}L)}\\
\end{bmatrix} 
\end{align*}

\noindent (c) clamped beam :
\begin{align*}
[F(\omega)]=
\begin{bmatrix}
1 & 0 & 1 & 1 & 1 & 1& 1 & 1\\
1 & L & {e}^{(m_{1}L)} & {e}^{(m_{2}L)} & {e}^{(m_{3}L)} & {e}^{(n_{1}L)}& {e}^{(n_{2}L)} & {e}^{(n_{3}L)} \\
0 & 1 & {m_{1}} & {m_{2}} & {m_{3}} & {n_{1}}& {n_{2}} & {n_{3}}\\ 
0 & 1 & {m_{1}}{e}^{(m_{1}L)} & {m_{2}}{e}^{(m_{2}L)} & {m_{3}}{e}^{(m_{3}L)} & {n_{1}}{e}^{(n_{1}L)}& {n_{2}}{e}^{(n_{2}L)} & {n_{3}}{e}^{(n_{3}L)} \\
0 & 0 & {m_{1}}^2 & {m_{2}}^2 & {m_{3}}^2 & {n_{1}}^2& {n_{2}}^2 & {n_{3}}^2\\ 
0 & 0 & m_{1}^{2}{e}^{(m_{1}L)} & m_{2}^{2}{e}^{(m_{2}L)} & m_{3}^{2}{e}^{(m_{3}L)} & n_{1}^{2}{e}^{(n_{1}L)} &  n_{2}^{2}{e}^{(n_{2}L)} & n_{3}^{2}{e}^{(n_{3}L)}\\
0 & 0 & {m_{1}}^3 & {m_{2}}^3 & {m_{3}}^3 & {n_{1}}^3& {n_{2}}^3 & {n_{3}}^3\\ 
0 & 0 & m_{1}^{3}{e}^{(m_{1}L)} & m_{2}^{3}{e}^{(m_{2}L)} & m_{3}^{3}{e}^{(m_{3}L)} & n_{1}^{3}{e}^{(n_{1}L)} &  n_{2}^{3}{e}^{(n_{2}L)} & n_{3}^{3}{e}^{(n_{3}L)}\\
\end{bmatrix} 
\end{align*}

\noindent (d) Propped cantilever beam :
\begin{align*}
[F(\omega)]=
\begin{bmatrix}
1 & 0 & 1 & 1 & 1 & 1& 1 & 1\\
0 & 1 & {m_{1}} & {m_{2}} & {m_{3}} & {n_{1}}& {n_{2}} & {n_{3}}\\ 
0 & 0 & {m_{1}}^2 & {m_{2}}^2 & {m_{3}}^2 & {n_{1}}^2& {n_{2}}^2 & {n_{3}}^2\\ 
0 & 0 & {m_{1}}^3 & {m_{2}}^3 & {m_{3}}^3 & {n_{1}}^3& {n_{2}}^3 & {n_{3}}^3\\ 
1 & L & {e}^{(m_{1}L)} & {e}^{(m_{2}L)} & {e}^{(m_{3}L)} & {e}^{(n_{1}L)}& {e}^{(n_{2}L)} & {e}^{(n_{3}L)} \\
0 & 0 & m_{1}^{2}{e}^{(m_{1}L)} & m_{2}^{2}{e}^{(m_{2}L)} & m_{3}^{2}{e}^{(m_{3}L)} & n_{1}^{2}{e}^{(n_{1}L)} &  n_{2}^{2}{e}^{(n_{2}L)} & n_{3}^{2}{e}^{(n_{3}L)}\\
0 & 0 & t_{3}{e}^{(m_{1}L)}&t_{4}{e}^{(m_{2}L)}&t_{5}{e}^{(m_{3}L)}&t_{6}{e}^{(n_{1}L)}&t_{7}{e}^{(n_{2}L)}&t_{8}{e}^{(n_{3}L)}\\
0 & 0 & m_{1}^{3}{e}^{(m_{1}L)} & m_{2}^{3}{e}^{(m_{2}L)} & m_{3}^{3}{e}^{(m_{3}L)} & n_{1}^{3}{e}^{(n_{1}L)} &  n_{2}^{3}{e}^{(n_{2}L)} & n_{3}^{3}{e}^{(n_{3}L)}\\
\end{bmatrix} 
\end{align*}

\end{document}